\documentclass{aa}

\usepackage{natbib}
\usepackage{txfonts}
\usepackage[dvips]{graphicx}
\bibpunct{(}{)}{;}{a}{}{,}
\usepackage{aalongtable}

\begin{document}

\title{Spectral Energy Distributions of a Large Sample of BL Lacertae Objects}

\author{E. Nieppola\thanks{e-mail: eni@kurp.hut.fi} \inst{1}, M. Tornikoski \inst{2} and E. Valtaoja
       \inst{1,3}  
       }

\institute{Tuorla Observatory, V\"ais\"al\"antie 20, 
		FIN--21500 Piikki\"o, Finland
		\and 
		Mets\"ahovi Radio Observatory, Mets\"ahovintie 114,
		FIN--02540 Kylm\"al\"a, Finland
		\and
		Dept. of Physical Sciences, University of Turku, FIN--20100
		Turku, Finland\\
		}

\date{Received / Accepted}


\abstract{We have collected a large amount of multifrequency data for the objects in the
  Mets\"ahovi Radio Observatory BL Lacertae sample and computed their spectral
  energy distributions (SED) in the log\,$\nu$ -- log\,$\
\nu F$ --\,representation. This is the first
  time the SEDs of BL Lacs have been studied with a sample
  of over 300 objects. The synchrotron components of the SEDs were fitted with a parabolic function
  to determine the synchrotron peak frequency, $\nu_{peak}$. We checked the
  dependence between luminosities at several frequency bands and synchrotron peak frequency to
  test the blazar sequence scenario, which states that the source luminosity
  depends on the location of the synchrotron peak. We also calculated broad
  band spectral indices and plotted them against each other and $\nu_{peak}$.

  The range of $\nu_{peak}$ in our study was considerably extended compared to
  previous studies. There were 22 objects for which
  log\,$\nu_{peak}>$19. The data shows that at 5 GHz, 37 GHz and 5500 $\textrm{\AA}$ there is negative
  correlation between luminosity and $\nu_{peak}$, whereas in X-rays the
  correlation turns slightly positive. There is no significant correlation
  between source luminosity at synchrotron peak and $\nu_{peak}$. Several low radio
  luminosity--low energy peaked BL Lacs were found. The negative correlation
  between broad band spectral indices and $\nu_{peak}$ is also significant,
  although there is substantial scatter. Therefore we find that neither $\alpha_{rx}$
  nor $\alpha_{ro}$
  can be used to determine the synchrotron peak of BL Lacs. On the grounds of our results we
  conclude that the blazar sequence scenario is not valid. In all our results
  the BL Lac population is continuous with no hint of the bimodality of the
  first BL Lac samples.


}

\authorrunning{E. Nieppola et al.}
\titlerunning{SEDs of BL Lacertae Objects}

\maketitle


\section{Introduction}

BL Lacertae (BL Lacs) are a subclass of active galactic nuclei (AGN). They are
characterized by the lack of strong emission lines, rapid variability at all wavelenghts
and strong polarization. Their spectral energy distribution (SED), in the log\,$\nu$ -- log\,$\nu F$ --\,representation, consists of
a synchrotron component at lower frequencies and an inverse Compton component at higher
frequencies. The peculiar traits of the BL Lac class are most likely caused by
Doppler--boosted radiation emanating from a relativistic jet aligned close to the line
of sight \citep{urry95}. 

Traditionally BL Lacs have been discovered in either
radio or X-ray band, which led to their classification as radio-selected
(RBL) and X-ray-selected (XBL) BL Lacs. Best-known RBL samples include the 1Jy
\citep{stickel91}, S4 \citep{stickel94_S4} and S5 \citep{kuhr90} samples and
among the most important XBL
samples are the EMSS \citep{gioia90,stocke91} and Slew Survey
\citep{perlman96}.

 The two classes have different
properties: RBLs are more variable, more luminous at radio and optical
wavelengths and have a higher polarization
\citep{stocke85,jannuzi94}. XBLs have a higher starlight fraction, 30--50\,\%
\citep{stocke85}, and their morphology is less core--dominated in the radio
than that of RBLs \citep{perlman93}.
Due to these differences they were initially
regarded as separate classes of AGN. However, in recent years
samples including intermediate objects have been found in surveys which
combine X-ray and radio observations. These include the
RGB (RASS--Green Bank) sample \citep{laumueh99}, the Deep X-Ray Radio Blazar
Survey (DXRBS) \citep{perlman98,landt01} and the REX Survey \citep{caccianiga99}. Their
discovery has strengthened the view that the BL Lac population is
continuous and RBLs and XBLs represent the opposite ends of the continuum. The reason for such a continuity lies in the cutoff frequency of the
synchrotron component in the SED. The
synchrotron peak frequency of RBLs is in the radio/IR band and for XBLs the
peak is mostly in
the UV/X-ray band \citep{giommi95}. The intermediate BL Lacs (IBL) have
their synchrotron peak in the optical wavelengths. This explains why they were
not observed in the first surveys. Following the synchrotron cutoff model the
terminology was brought to more physical ground: the
RBL\,/\,XBL\,-division was replaced by division into low-energy-peaked BL Lacs
(LBL) and high-energy-peaked BL Lacs (HBL) \citep{padovani95}. Most
RBLs are LBLs and most XBLs are HBLs, but not all. The class boundaries can be
loosely defined as $\nu_{peak} \approx 10^{13-14}$ Hz for LBLs, $\nu_{peak}
\approx 10^{15-16}$ Hz for IBLs and $\nu_{peak} \approx 10^{17-18}$ Hz for
HBLs.

\citet{fossati98} linked the shape of the SED and the synchrotron peak
frequency to the source luminosity: the lower the peak frequency, the more
luminous the source. This would mean that LBLs are intrinsically more luminous
than HBLs. This sequencing is based on the absence of high-luminosity HBLs and
low-luminosity LBLs. However, recently \citet{giommi05} reported the possible
discovery of high-luminosity HBLs in the Sedentary Survey and evidence of low-power
LBLs has also been discovered \citep{padovani03,caccianiga04}. These findings are at odds with the
trend presented by Fossati et al.

\citet{ghisellini99} suggested that there is a class of BL Lacs whose synchrotron peak
lies at even higher frequencies than that of conventional HBLs, $\nu_{peak}\,>\,10^{19}$ Hz. These objects can be called
ultra-high-energy synchrotron peak BL Lacs (UHBLs) \citep{giommi01}. Following the
dependency of the SED shape and luminosity, UHBLs are thought to be extremely faint at radio
wavelengths which is why they have escaped notice. Extensive $\gamma$-ray
observations are needed to unambiguously identify them. 

In this paper we take a new approach to studying the properties of the BL Lac
population. Our goal is to plot the SEDs of the Mets\"ahovi Radio Observatory BL Lac sample, which
comprises nearly 400 objects, including objects from all the best-known
surveys at radio and X-ray wavelengths. This way we can examine the population
properties of a sample with a wide range of attributes instead of focusing on
one or two limited surveys.

The aim of this paper was to test both the continuity of the BL Lac population and
the blazar sequence scenario, and to assign a SED-based classification to those objects that previously had none. A large
database of flux measurements is collected and SEDs are plotted for more than
300 BL Lacs. Each object is classified as LBL, IBL or HBL. The relationship of
synchrotron peak frequency and luminosities at several frequencies is also
tested along with the properties of broad band spectral indices. Throughout
this paper we assume
$H_{0}=65\,\textrm{km}\,\textrm{s}^{-1}\,\textrm{Mpc}^{-1}$ and
$\Omega_{0}=1$. All
statistical tests were carried out using Unistat 5.5 software.

\section{The sample}

The Mets\"ahovi BL Lac sample includes 381 objects selected from the Veron-Cetty
\& Veron BL Lac Catalogue \citep{veron00}, hereafter VCV2000, and 17 objects from literature. Given the
northern location of the Mets\"ahovi observatory, the source with the lowest
declination in the sample is PKS 2223-114 at $\delta=$\,--11:13:41. The list
of sample sources (table~\ref{sample}) is published electronically\footnote{www.edpsciences.org}. A large part
of the
objects in VCV2000 are from well-known and well-defined samples such as the 1Jy, S4,
S5, EMSS and Einstein Slew Survey. Also BL Lacs from the first release of DXRBS identifications
\citep{perlman98} are included. No selection criteria (other than declination)
in addition to the ones
in the original surveys were imposed on the sample. The aim was to examine the
behaviour of an extensive sample containing all known BL Lacs up to the year
2000. 

According to VCV2000 the Mets\"ahovi BL Lac sample can be further classified as follows: 63
\% are confirmed, 3 \% probable and 8 \% possible BL Lacs, 14 \% are objects
with high optical polarization and 12 \% (including the BL Lacs taken from
literature) lack any subclassification. In the sample there are 6 sources that
are not in the later editions of Veron-Cetty \& Veron BL Lac Catalogues
\citep{veron01,veron03}. These objects have been excluded from the data
analysis performed in this paper.

\section{The data}

In order to plot as accurate SEDs as possible, a large amount of
data from several wavelengths was collected. Because simultaneous
multifrequency flux measurements are not available, datapoints from different
epochs were searched
from databases and literature.

\subsection{Radio data}

The starting point in collecting radio data were the Mets\"ahovi observations at
37 GHz from
late 2001 to January 2004. The full data set and a more detailed analysis
about the 37 GHz behaviour of the sample sources will be published in a
forthcoming paper (Nieppola et al., in preparation for the A\&A). Of the BL
Lac sample 137 objects were detected at
$\sigma\geq4$. There were 255 BL Lacs that were not detected and 6 that had
not been observed yet during the time mentioned. The limiting flux of the
Mets\"ahovi radio telescope is about 0.2 Jy under optimal weather conditions.

Our group has also obtained flux density data at higher radio frequencies from
our observations with the Swedish--ESO Submillimetre Telescope (SEST) between
1987 and 2003 at 3 mm and 1.3 mm. For some objects we got data from the RATAN
observatory at frequencies 2.3, 4.8, 7.7, 11.2, 21.8 and 30 GHz (Tornikoski et al., in preparation for the A\&A). Additional low
frequency datapoints were found in VCV2000 and WGA--catalogue \citep{white96}.

The large majority of radio data were obtained from the Astrophysical
Catalogues Support System (CATS) maintained by Special Astrophysical
Observatory, Russia\footnote{http://cats.sao.ru/.}. The search results from CATS contained data
from over 140 different catalogues and more than 100 radio frequencies. 

\subsection{Data from other frequency bands}

The IR datapoints are from CATS, originating from IRAS- and
2MASS--catalogues. The wavelengths used are 1.25, 1.65, 2.0, 2.17, 12, 25, 60
and 100 $\mu$m. 

The optical data also are mainly from CATS. Some datapoints from V--band were added from \citet{donato01}.

The X-ray data are from Einstein- and ROSAT--catalogues. The majority is from
WGACAT \citep{white96} and RBSC--catalogue \citep{voges99}. Datapoints were
also collected from the following papers: \citet{donato01}, \citet{lamer96}
and \citet{laumueh99}. EMSS--data were included as well \citep{gioia90,stocke91}.

All of the data from gamma region came from the Third EGRET Catalogue
\citep{hartman99_3EG}. Such data were available for only 14 objects.

\section{Computing the spectral energy distributions}

For all sources with a sufficient number of datapoints a spectral energy
distribution was plotted in the form log\,$\nu$--log\,$\nu F$. All frequencies
used are observed frequencies, they have not been reduced to rest frame frequencies. The synchrotron
component of the SED was
fitted with a parabolic function \begin{equation}y=Ax^2+Bx+C\end{equation} in order to determine the synchrotron peak
frequency $\nu_{peak}=-B/2A$. The fitting was successful for 304 objects; the
rest were too sparsely sampled. The decision whether or not to include X-ray
datapoints in the fit was based solely on a visual estimate for each
individual object. All the SEDs (fig.~\ref{sed1}) are published electronically. The synchrotron peak frequencies are shown in table~\ref{sample}
col. 5.

We note that using a simple parabolic function in the fitting
produces some error especially among HBLs. In their case the X-ray datapoints
are typically included in the rising synchrotron component, and therefore the
parabola peaks after the X-ray domain. In reality, the synchrotron peak is
expected to occur at or very close to the X-ray wavelengths and the decline to
be more rapid. Thus the peak frequencies of the most extreme objects can be
exaggerated.

The objects were assigned a LBL/IBL/HBL classification according to
$\nu_{peak}$. On the basis of the criteria of \citet{padovani95}, we decided
to draw the boundaries as follows: for LBLs, log\,$\nu_{peak}<$\,14.5, for IBLs
14.5\,$<\,$log\,$\nu_{peak}<$\,16.5 and for HBLs
log\,$\nu_{peak}>$\,16.5. Thus the peak frequencies of LBLs stretch up to the
optical region, IBLs peak in the optical and UV--bands and HBLs from soft
X-rays upwards. 

\subsection{The distribution of $\nu_{peak}$}

\begin{figure}
\resizebox{\hsize}{!}{\includegraphics{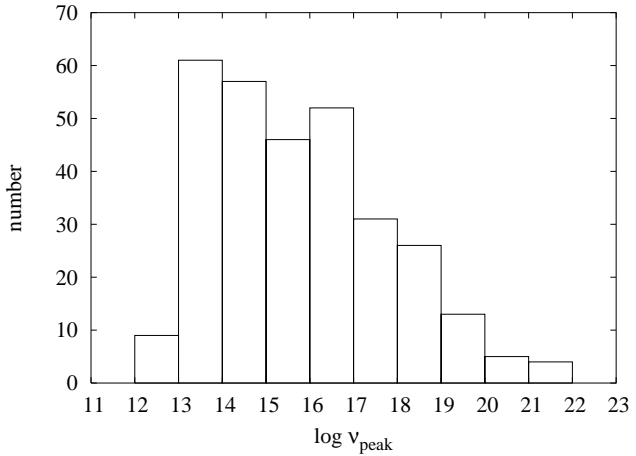}}
\caption{The distribution of the synchrotron peak frequencies in the
  Mets\"ahovi sample.}
\label{distribution}
\end{figure}

When the objects were classified as LBL/IBL/HBL according to their $\nu_{peak}$ as
described in the previous section, the division resulted in the three classes
being almost equal in size. There were 98
LBLs, 96 IBLs and 110 HBLs. The distribution is smooth and decreases steadily
towards the higher peak frequencies (fig.~\ref{distribution}). This can be a real effect, suggesting that
sources in which electrons are accelerated to extreme energies are
intrinsically rare, or due to selection effects in surveys. The frequency interval
$\nu_{peak}=10^{13-14}$ Hz is the most populated. 

Table \ref{division} presents how the classification with respect to the observational band
relates to the one based on the SED. Here RBL classification has been assigned
to objects in the 1Jy and S4 surveys, IBL to objects in RGB and 200 mJy BL Lac \citep{bondi01}
surveys and XBL to objects in the Einstein Slew Survey. This RBL/XBL
classification has been adopted from \citet{giommi95}. Among the 304 sources
for which the SED could be plotted there were 31 RBLs, 115
IBLs and 48 XBLs. Several objects got multiple classifications.

\begin{table}
\caption{Division of observational BL Lac classes to physical ones.}
\label{division}
\centering
\begin{tabular}{c c c c c}
\hline\hline
 & \vline & LBL & IBL & HBL\\
\hline\hline
RBL & \vline & 84 \% & 10 \% & 6 \% \\
IBL & \vline & 22 \% & 36 \% & 42 \% \\
XBL & \vline & 21 \% & 25 \% & 54 \% \\
\hline\hline
\end{tabular}
\end{table}

Table \ref{division} clearly demonstrates how surveys in the X-ray energies are more prone
to find low-energy BL Lacs than radio surveys are to find HBLs. A large
majority of RBLs really are LBLs. In fact, the 6 \% of RBLs that turned out to
be HBLs
are the radio--luminous Mrk 421 and Mrk 501. Objects in the RGB and 200 mJy samples are
more likely to be HBLs than LBLs, and only a third of them are truly
intermediate. Of XBLs only half are HBLs and over 20 \% are actually low-energy
BL Lacs. These figures are certainly affected by the arbitrariness of the
dividing boundaries between the classes, but the overall trend is expected to
remain. It results from the fact that the X-ray luminosities of the samples in question
are roughly the same whereas radio luminosities differ greatly. 

\subsection{UHBL candidates}

As fig.~\ref{distribution} shows, there were several objects in the sample whose synchrotron
peak frequency was extremely high. Usually objects with log\,$\nu_{peak}>$\,17
are considered as extreme, here the number of such objects was 80,
approximately 26 \% of
all the fits. For 22 objects log\,$\nu_{peak}>$\,19 (table \ref{UHBL}) and for 9 objects even
log\,$\nu_{peak}>$\,20, corresponding to a peak energy of $\sim$\,0.4 MeV. The SEDs of these sources are
mainly very sparsely sampled, typically with datapoints from radio, optical
and X-ray bands, and should be treated with caution. With that said, we note that
even for objects with log\,$\nu_{peak}>$\,20 the datapoints fit very well on
the rising parabolic function. We note again that the actual position of the
peak is probably exaggerated by the use of a parabolic fitting function as
mentioned in sect. 4. Therefore the peak frequencies of these objects cannot
be considered as definite. 
Peaking near the
MeV--region these sources would be excellent candidates for
$\gamma$--observations.

\begin{table*}
\caption{Objects for which log\,$\nu_{peak}>$19.}
\label{UHBL}
\centering
\begin{tabular}{l c c c}
\hline\hline
Source & RA. (J2000) & Dec. (J2000) & log\,$\nu_{peak}$\\
\hline\hline
1ES 0229+200 & 02:32:48.6 & +20:17:17 & 19.45\\
RXS J0314.3+0620 & 03:14:23.9 & +06:19:57 & 19.57\\
2E 0323+0214 & 03:26:13.9 & +02:25:14 & 19.87\\
2E 0414+0057 & 04:16:52.4 & +01:05:24 & 20.71\\
1ES 0502+675 & 05:07:56.1 & +67:37:24 & 19.18\\
EXO 0706.1+5913 & 07:10:30.1 & +59:08:21 & 21.05\\
RXS J0847.2+1133 & 08:47:12.9 & +11:33:52 & 19.13\\
1ES 0927+500 & 09:30:37.5 & +49:50:25 & 21.13\\
RXS J1008.1+4705 & 10:08:11.3 & +47:05:20 & 19.67\\
RXS J1012.7+4229 & 10:12:44.3 & +42:29:57 & 20.97\\
EXO 1449.9+2455 & 11:49:30.3 & +24:39:27 & 19.83\\
PG 1218+304 & 12:21:21.9 & +30:10:37 & 19.14\\
RXS J1319.5+1405 & 13:19:31.7 & +14:05:34 & 20.85\\
RXS J1341.0+3959 & 13:41:05 & +39:59:45 & 20.06\\
RXS J1353.4+5601 & 13:53:28 & +56:00:55 & 19.23\\
RXS J1410.5+6100 & 14:10:31.7 & +61:00:10 & 20.25\\
2E 1415+2557 & 14:17:56.6 & +25:43:25 & 19.24\\
RXS J1456.0+5048 & 14:56:03.7 & +50:48:25 & 19.94\\
RXS J1458.4+4832 & 14:58:28 & +48:32:40 & 21.46\\
1ES 1533+535 & 15:35:00.8 & +53:20:37 & 19.68\\
RXS J1756.2+5522 & 17:56:15.9 & +55:22:18 & 19.90\\
RXS J2304.6+3705 & 23:04:36.6 & +37:05:08 & 21.01\\
\hline\hline
\end{tabular}
\end{table*}

\section{Correlation between $\nu_{peak}$ and luminosity}

\subsection{High radio frequency (37 GHz)}

Out of the three BL Lac classes, LBLs had 37 GHz detections for
  81 \% of the sources, IBLs for 36 \% and HBLs for 12
\%. This seems to indicate that most HBLs have a radio flux well below the flux limit of the Mets\"ahovi telescope. This prompted us to study the possible correlation between the synchrotron peak frequency and 37 GHz source luminosity more closely.

To calculate the luminosities, redshift data was collected from VCV2000, \citet{stocke91}, \citet{laumueh99},
\citet{donato01} and SIMBAD
database\footnote{http://simbad.u-strasbg.fr.}. For sources which had no
redshift available we used $z$\,=\,0.4. Both a detection at 37 GHz and
synchrotron peak frequency $\nu_{peak}$ were available for 132 objects. 

\begin{figure}
\resizebox{\hsize}{!}{\includegraphics{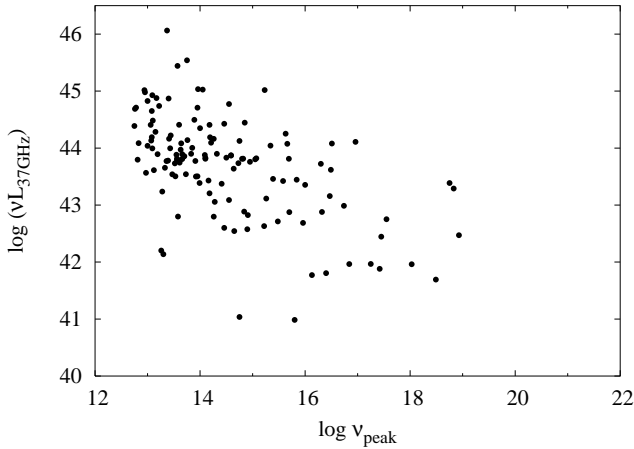}}
\caption{Radio luminosity at 37 GHz plotted against synchrotron peak
  frequency.}
\label{37GHz}
\end{figure}

When the 37 GHz luminosity is plotted against the synchrotron peak frequency (fig.~\ref{37GHz}), the
correlation is easily seen. LBLs are at the high-luminosity end of the plot
and the luminosity drops towards the high-energy regime. According to Spearman
Rank Correlation Test, there is significant negative correlation at the 99 \%
confidence level.  

No real evidence of a population of
low-luminosity LBLs or high-luminosity HBLs is found. However, there are two
LBLs with almost as low luminosities as HBLs disrupting the declining
trend. In addition, the lowest luminosities are not claimed by objects in the high end of
the peak frequency range, but by two IBLs with log\,$\nu_{peak}<$\,16. The radio luminosities are widely
scattered. An object with log\,$\nu L$\,=\,44 can have peak frequency values
ranging from
log\,$\nu_{peak}$\,=\,13 to log\,$\nu_{peak}$\,=\,17, approximately. Thus the radio
luminosity cannot be used to determine the peak frequency of the source.

\subsection{Low radio frequency (5 GHz)}

\begin{figure}
\resizebox{\hsize}{!}{\includegraphics{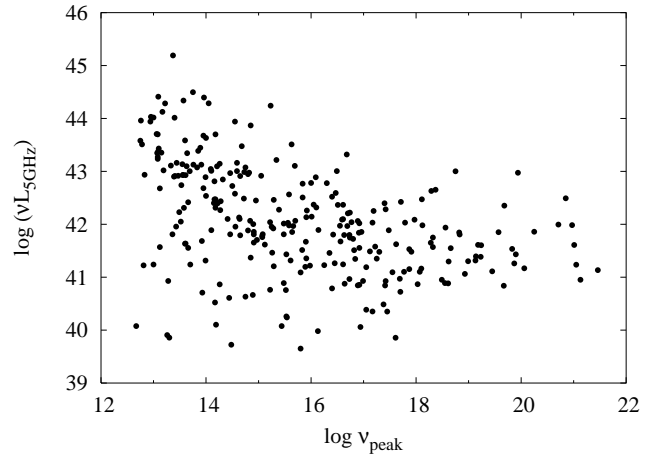}}
\caption{Radio luminosity at 5 GHz plotted against synchrotron peak
  frequency.}
\label{5GHz}
\end{figure}

The small number of HBL datapoints at 37 GHz convinced us to test
the correlation also at 5 GHz. The number of available datapoints rose to 280. The
correlation plot changed drastically (fig.~\ref{5GHz}). The most noticeable difference
is the appearance of several low-luminosity LBLs (lower left of the figure). They even reach lower
luminosities than any of the HBLs. The overall negative correlation is still present and significant at 99 \% level. This differs from the result obtained by
\citet{padovani03}. However, their DXRBS BL Lac sample consisted of only 31
objects mainly representing the LBL/IBL end of the plot. This is the region with
most scatter in our figure and within this limited area the correlation is less
obvious even with a larger number of datapoints. Only when the whole range of values of log\,$\nu_{peak}$ is
considered the trend becomes evident.

When compared with the corresponding figure of \citet{fossati98} our figure
has much more scatter. A part of it is caused by a larger number of datapoints,
but on closer examination our log\,$\nu_{peak}$ values for the Slew
Survey are much more widely spread. In \citet{fossati98} the Slew Survey
objects have log\,$\nu_{peak}$\,=\,15-19, while in our study they have
log\,$\nu_{peak}\approx$\,13-21. The 1Jy--sample takes on similar values of
peak frequency in both studies.

We note that there is no evidence of very high-luminosity HBLs. In fact, the
extreme HBLs adopt quite intermediate luminosity values avoiding also the
low-luminosity region. However, the spectra of possible high-luminosity HBLs would
  very likely be totally featureless because of the powerful nucleus and thus the object would lack a
  redshift estimation. Therefore the redshift value assigned to featureless
  sources could have a big impact on the luminosity correlations and the appearance of
  high-luminosity HBLs (see section 5.6). 

\subsection{Optical region (5500 \AA)}

\begin{figure}
\resizebox{\hsize}{!}{\includegraphics{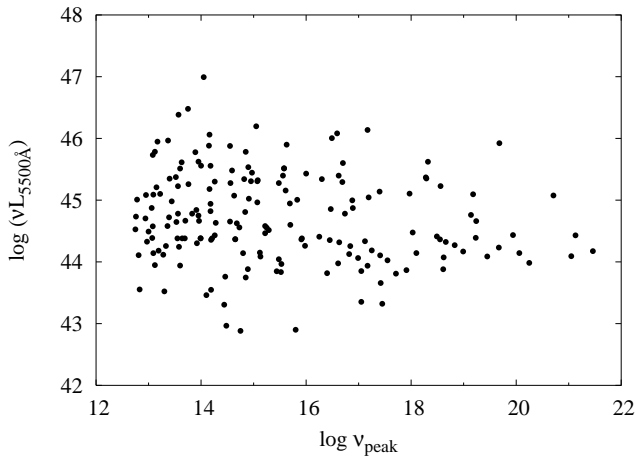}}
\caption{Optical luminosity at 5500 $\textrm{\AA}$ plotted against synchrotron peak
  frequency.}
\label{5500}
\end{figure}

Figure~\ref{5500} shows the optical luminosity at wavelength 5500 $\textrm{\AA}$ plotted against
log\,$\nu_{peak}$. There is a significant, slightly negative correlation
present at 95 \% level. We see again that LBLs have more scatter than
HBLs. Overall, the correlation is much less evident than in the case of radio
luminosity. 

\subsection{X-ray region}

\begin{figure}
\resizebox{\hsize}{!}{\includegraphics{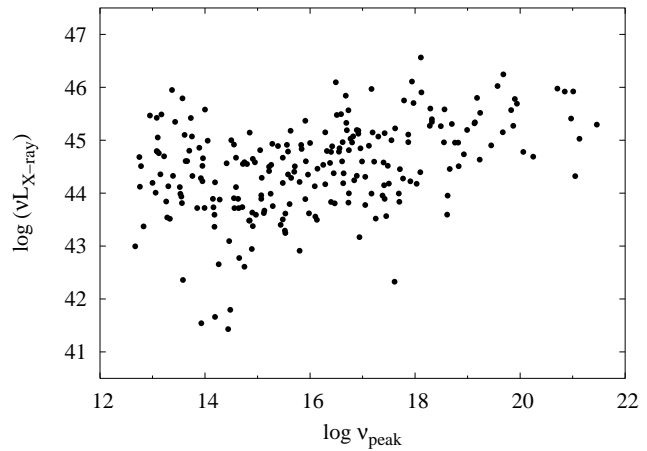}}
\caption{X-ray luminosity at 1 keV and ROSAT band plotted against synchrotron peak
  frequency.}
\label{1keV}
\end{figure}

In fig.~\ref{1keV} we have plotted the correlation between X-ray luminosity
and log\,$\nu_{peak}$. In calculating the luminosities we have used both 1 keV
data and data from the ROSAT band (0.1--2.4 keV). The error produced by the
bandwidth difference is negligible when only the statistical properties of a
large sample are considered. We note that in the case of X-ray luminosity, the correlation is positive
and significant at 99 \% level. This is in contrast with the blazar
sequence scenario. \citet{fossati98} state that the overall luminosity of
HBLs is lower than that of LBLs (see their fig. 12). While they admit that
in the X-ray band objects exhibit complex behaviour, the systematic rising
trend presented by our findings is not predicted. 

\subsection{Peak luminosity $L_{peak}$}

\begin{figure}
\resizebox{\hsize}{!}{\includegraphics{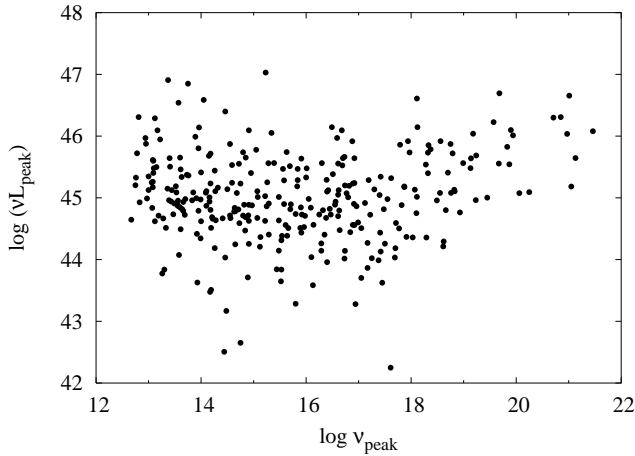}}
\caption{Luminosity at the synchrotron peak frequency plotted against synchrotron
  peak
  frequency.}
\label{peaklum}
\end{figure}

In addition to luminosities at defined frequency bands, we calculated for each
source its luminosity at the synchrotron peak frequency. This is plotted
against log\,$\nu_{peak}$ in fig.~\ref{peaklum}. There is no significant
correlation. Therefore we can decisively say that the source luminosity does
not depend on the synchrotron peak frequency. Fig. 7 of \citet{fossati98} also
shows the dependence of $L_{peak}$ and log\,$\nu_{peak}$. In their study
there is a significant, yet weak correlation. Again, we remark that a larger
number of datapoints and a wider range of log\,$\nu_{peak}$ reveal the true
behaviour of the population and the lack of correlation. If only objects with
log\,$\nu_{peak}<\,$17 are considered, there is a weak negative correlation also
in our sample. On the other hand, when the high-energy tail with
log\,$\nu_{peak}>\,$17 is tested, we find a significant positive
correlation. Therefore the distribution almost seems to assume a concave shape.

\subsection{The effect of $z$ on luminosity correlations}

\citet{giommi05} found numerous candidates for high-luminosity HBLs in the
Sedentary Survey. All these objects seemed to reside at high redshifts
($z\geq$\,0.7). In our luminosity calculations we assumed $z$\,=\,0.4 for
featureless objects, which is a low value compared to Giommi et al. Using
a too low redshift value for a significant part of the sources would lead to a
serious underestimation of luminosities. To
take this into account, we tested the luminosity correlations also by assigning
redshifts $z$\,=\,0.7, $z$\,=\,1 and $z$\,=\,1.5 to those objects that had none. 

The effect was most pronounced in the case of radio luminosities at 5 and 37
GHz. The number of relatively high-luminosity HBLs increased with the
higher assumed redshift. However, only at very high reshift values ($z$\,=\,1
or $z$\,=\,1.5) the luminosities of a few HBLs became roughly comparable to those of
LBLs. Considering the population average, $z_{av}$\,=\,0.33, it is questionable whether all featureless sources would have $z\,\geq$\,1, although for some of them this may be the case. Therefore,
while the existence of high-luminosity HBLs in this sample is possible, we do not expect the correlations to be affected by them.

In other frequency bands than radio, the effect of an increasing redshift was
negligible. The shape of the correlations did not change notably, only the scatter increased somewhat. We note that in all wavelengths and for all assumed
redshift values the significance of the
statistical correlations remained the same.

\section{Other properties of the sample}

\subsection{Log\,$(S_{x}/S_{r})$\,--distribution}

The log\,$(S_{x}/S_{r})$--distribution has traditionally been used to point
out the bimodality in the BL Lac population. The dividing line between RBLs
and XBLs has been
log\,$(S_{x}/S_{r})$=$-5.5$ \citep{laumueh99}, when the fluxes are in the same
units and X-ray and radio frequencies are 0.1--2.4 keV and 5 GHz
respectively. Here we calculated the distribution to check the assumed
continuity of the sample. X-ray fluxes are from the ROSAT band 0.1-2.4 keV and
radio measurements are from Mets\"ahovi at 37 GHz. If there was more than one
flux measurement from one frequency, the average value was used. 

\begin{figure}
\resizebox{\hsize}{!}{\includegraphics{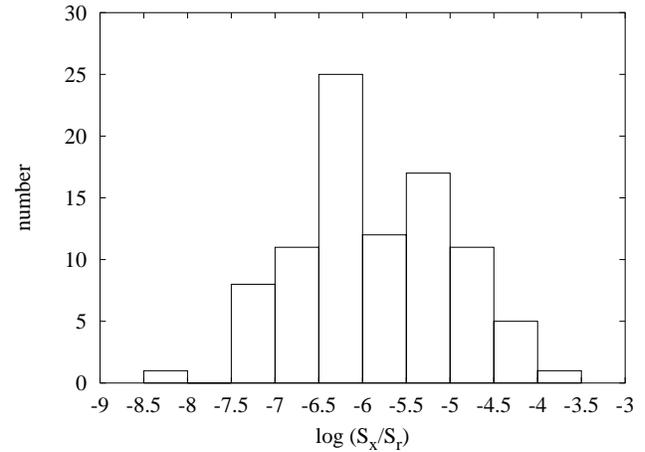}}
\caption{The distribution of log\,$(S_{x}/S_{r})$.}
\label{distribution2}
\end{figure}

The distribution is indeed unbroken with an average of
log\,$(S_{x}/S_{r})$=$-5.85$ (fig.~\ref{distribution2}). The most populated interval is
log\,$(S_{x}/S_{r})$=[$-6$,\,$-6.5$]. Pertaining to the unbalanced detection rates at 37
GHz, LBLs are overrepresented compared to HBLs in the plot. Only 14 \% of the
objects are HBLs. When considered separately, LBLs, IBLs and HBLs move
progressively from low to higher values of log\,$(S_{x}/S_{r})$. There is
substantial overlap between the distributions. The average values are $-$6.35,
$-$5.50 and $-$4.85 for LBLs, IBLs and HBLs respectively.

\subsection{Broad band spectral indices}

\subsubsection{The $\alpha_{ro}-\alpha_{ox}$ diagram}

We calculated broad band spectral indices $\alpha_{ro}$ and $\alpha_{ox}$ to
plot the sample in $\alpha_{ro}$ vs. $\alpha_{ox}$ diagram. Here the
spectral index is defined as
\begin{equation}\alpha_{1-2}=-\frac{log(S_{1}/S_{2})}{log(\nu_{1}/\nu_{2})}\end{equation}
where $S_{1}$ and $S_{2}$ are the fluxes in frequencies $\nu_{1}$ and
$\nu_{2}$ respectively and $S_{\nu}=\nu^{-\alpha}$. 

The $\alpha_{ro}$ vs. $\alpha_{ox}$ diagram has
also been used as a means to demonstrate the division of the
population. Generally, XBLs
have lower values of both indices occupying the lower left corner of the
plot, whereas RBLs lie in the upper right
corner \citep{stocke85,laumueh99}. IBLs seem to have bridged
the gap having intermediate values of both indices.

\begin{figure}
\resizebox{\hsize}{!}{\includegraphics{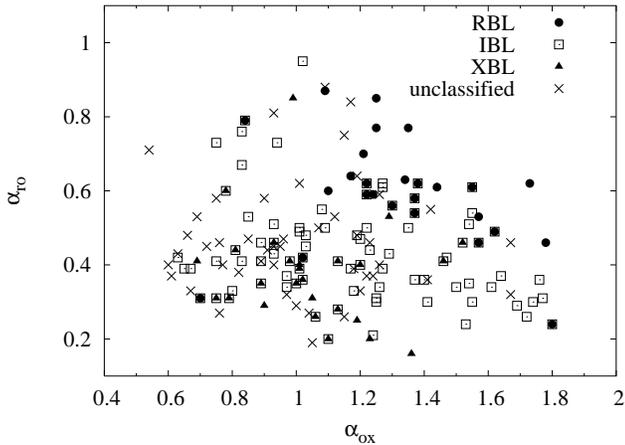}}
\caption{The colour plane: $\alpha_{ro}$ plotted against
  $\alpha_{ox}$. Frequency intervals 5 GHz--5500 \AA--1 keV, ROSAT
  band. Different symbols denote the observational classification of BL Lacs.}
\label{alallow}
\end{figure}

\begin{figure}
\resizebox{\hsize}{!}{\includegraphics{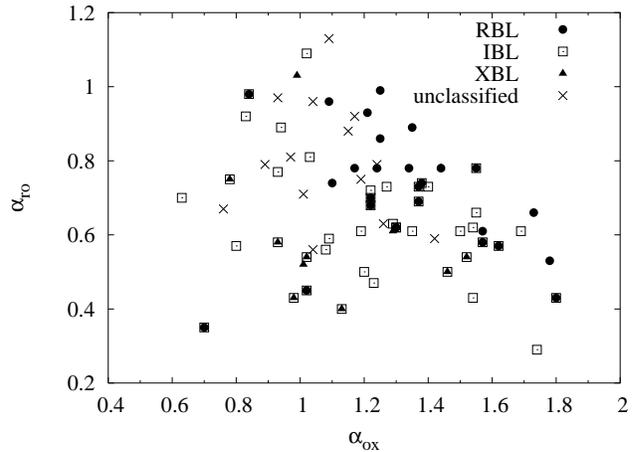}}
\caption{The colour plane: $\alpha_{ro}$ plotted against
  $\alpha_{ox}$. Frequency intervals 37 GHz--5500 \AA--1 keV, ROSAT
  band. Different symbols denote the observational classification of BL Lacs.}
\label{alalhigh}
\end{figure}

For comparison, we plotted two diagrams; one with low radio frequency (5 GHz) (fig.~\ref{alallow})
and one with high radio frequency (37 GHz) (fig.~\ref{alalhigh}). The optical wavelength used was
5500 \AA. In both diagrams we used X-ray data at 1 keV from literature and, in
addition, data from ROSAT band 0.1--2.4 keV. The error produced by the bandwidth
difference is small compared with the benefits of a larger number of
datapoints.  

When the lower radio frequency is used (fig.~\ref{alallow}), the indices for the whole
population are $\alpha_{ox}$\,=\,0.54--1.8 and $\alpha_{ro}$\,=\,0.16--0.95. RBLs and
XBLs occupy their expected locations but overall the distribution is even. 

As for the higher radio frequency diagram (fig.~\ref{alalhigh}), the intervals are slightly
different; $\alpha_{ox}$\,=\,0.63--1.8 and $\alpha_{ro}$\,=\,0.29--1.13. The values of
$\alpha_{ox}$ are the same as in fig.~\ref{alallow}, but the number of datapoints is
smaller. Here we note again the lack of HBLs compared to LBLs. This is why the
lower left corner of fig.~\ref{alalhigh} is underpopulated compared with fig.~\ref{alallow}. 

\begin{figure}
\resizebox{\hsize}{!}{\includegraphics{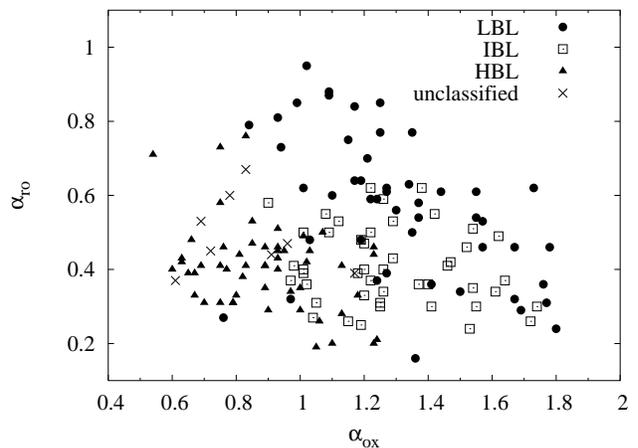}}
\caption{The colour plane: $\alpha_{ro}$ plotted against
  $\alpha_{ox}$. Frequency intervals 5 GHz--5500 \AA--1 keV, ROSAT
  band. Different symbols denote the physical classification of BL Lacs
  obtained in this study.}
\label{alalclass}
\end{figure}

We also plotted the diagram with 5 GHz showing the LBL/IBL/HBL classification
of our study (fig.~\ref{alalclass}). It clearly shows how the transition of the synchrotron
peak from LBL to HBL moves the object on the $\alpha_{ro}-\alpha_{ox}$ --plane from the top to
lower right and onwards to lower left, as described by
\citet{padovani95}. However, there are several LBLs that appear on the wrong
side of the $\alpha_{rx}$\,=\,0.75 line that has usually been thought of as a
dividing line between LBLs and HBLs. This suggests that the $\alpha_{rx}$\,=\,0.75
divide is not very effective and LBLs take on very scattered values of
$\alpha_{ro}$.
 
There are some objects in the $\alpha_{ro}-\alpha_{ox}$ --plot (fig.~\ref{alalclass}) with no
classification (marked with x). They are concentrated in quite a small area
with $\alpha_{ox}$\,=\,0.61--1.17 and $\alpha_{ro}$\,=\,0.37--0.67. Their synchrotron
peak frequencies have not been calculated on account of very poor fits, but
judging by their spectral indices they are likely to be HBLs. 

\subsubsection{$\alpha_{rx}$\,vs.\,log\,$\nu_{peak}$}

\begin{figure}
\resizebox{\hsize}{!}{\includegraphics{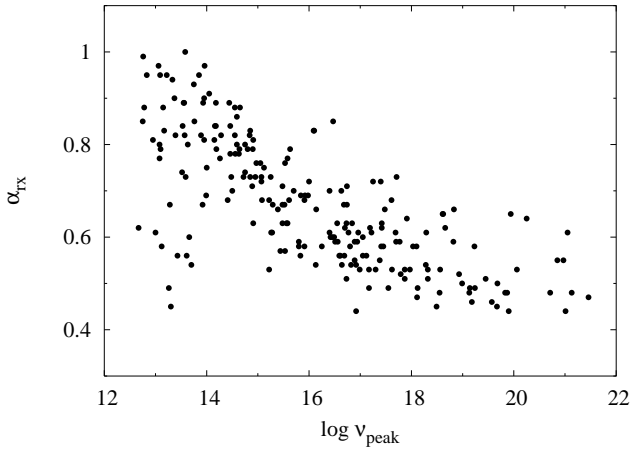}}
\caption{Radio-X-ray spectral index $\alpha_{rx}$ plotted against synchrotron
  peak frequency.}
\label{alrx}
\end{figure}

The relation between $\alpha_{rx}$ and log\,$\nu_{peak}$ is shown in
fig.~\ref{alrx}. We used radio frequency 5 GHz and X-ray data comes primarily
from 1 keV. For those objects that have no 1 keV flux available we used data
from the ROSAT band. According to the Spearman rank correlation test, the negative correlation is
significant at 99 \% level. However, the correlation seems to break apart at
log\,$\nu_{peak}<$\,14. LBLs take on $\alpha_{rx}$ --values between
0.45--1. \citet{fossati98} suggested that the correlation between the radio
luminosity and X-ray luminosity in the rising Compton--component makes
$\alpha_{rx}$ tend to a fixed value when log\,$\nu_{peak}<$\,14, but in our plot
this seems not to be the case. 

When log\,$\nu_{peak}>\,$14 the correlation is apparent. This speaks for the
authenticity of the UHBL candidates, as they take their place at the end of
the continuum with low spectral index values. All objects with log\,$\nu_{peak}>$\,19 also have $\alpha_{rx}\leq$\,0.65.

From the $\alpha_{rx}$\,vs.\,log\,$\nu_{peak}$ --plot we can see that
assigning an object an LBL/HBL classification based only on the value of
$\alpha_{rx}$ is risky. While object with $\alpha_{rx}>$\,0.75 is very likely an
LBL, the opposite does not hold. In fact, approximately 30 \% of objects with
$\alpha_{rx}<$\,0.75 have log\,$\nu_{peak}<$\,16. At $\alpha_{rx}\approx$\,0.6 the
possible values of $\nu_{peak}$ span across as much as eight magnitudes.

When compared with the corresponding figures of \citet{fossati98} and \citet{padovani03},
our plot seems to be a combination of the two. This is explained by the fact
that both the samples used by Fossati et al. (1Jy and Slew Survey) and part of
the sample used by Padovani et al. (DXRBS) are included in the Mets\"ahovi
sample. Thereby we see a strong correlation with a substantial increase in
scatter of the $\alpha_{rx}$--values towards the lower end. We stress that our
data considerably extends the range of values of log\,$\nu_{peak}$.

\subsubsection{$\alpha_{ro}$\,vs.\,log\,$\nu_{peak}$}

\begin{figure}
\resizebox{\hsize}{!}{\includegraphics{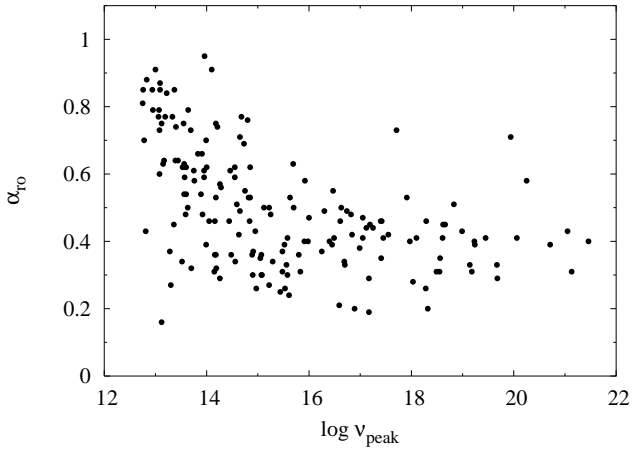}}
\caption{Radio-optical spectral index $\alpha_{ro}$ plotted against synchrotron
  peak
  frequency.}
\label{alro}
\end{figure}

The correlation between the spectral index between radio (5 GHz) and optical
(5500 \AA) frequencies and synchrotron peak frequency is shown in fig.~\ref{alro}. The overall trend is easily
seen, changing from the steep negative correlation of LBLs to an almost constant trend of
HBLs. The change occurs roughly when the synchrotron peak of the SED moves to
frequencies higher than those used to calculate $\alpha_{ro}$. After that
point, the peak frequency no longer has an influence on $\alpha_{ro}$. The
negative correlation is
significant at 99 \% level. Again, however, we
see a few points breaking this scenario, both LBLs with very low values of
$\alpha_{ro}$ and HBLs with high values of $\alpha_{ro}$.  

We note that, as in the case of $\alpha_{rx}$, the scatter in the figure is
significant. Therefore the value of $\alpha_{ro}$ cannot be reliably used to classify
BL Lacs. Only if $\alpha_{ro}$ is very close to one, the object is likely to
be an LBL. 

\section{Discussion}

Our findings support the results of \citet{costamante01}. Their study
concludes that objects 1ES
0033+595, 1ES 0120+340, 1ES 1218+304 and 1ES 1426+428 are extreme HBLs with
synchrotron peak frequencies exceeding $10^{18}$\,Hz. Our results show that they
all peak at roughly
log\,$\nu_{peak}\approx$\,19. Meanwhile, the extreme nature of 1ES 2344+514
\citep{giommi00} is not revealed in our study as it had a peak
frequency of log\,$\nu_{peak}$\,=\,16.4. Giommi et al. state that it is indeed
very variable, with synchrotron peak frequencies ranging from
log\,$\nu_{peak}\approx$\,17 to log\,$\nu_{peak}\approx$\,18. Mrk 501 is
another example of a variable source; its synchrotron peak has been as high as
log\,$\nu_{peak}\approx$\,19 \citep{pian98}, but the peak in our SED occured at log\,$\nu_{peak}\approx$\,16.84. 

In this paper we tested the correlation between the synchrotron peak frequency and
luminosities at radio, optical and X-ray wavelengths and luminosity at the
synchrotron peak of each source. According to the blazar sequence
scenario promoted by \citet{fossati98}, HBLs have lower luminosities at all
wavelenghts. In the X-ray region, in the SED of LBLs the synchrotron and
SSC--components meet resulting in a concave X-ray spectrum. In HBLs the
synchrotron component peaks at or near the X-ray energies. Because the SEDs
of LBLs and HBLs are so differently shaped in this frequency band, the X-ray correlation may be less
significant. Still, on the whole, LBLs are expected to be more luminous. When the
defined frequency bands are considered, our data seems to support this scenario in all
other frequencies  except in
X-rays. There we find a clear positive correlation which contradicts
the blazar sequence. Luminosities rise towards the high peak
frequencies. When all luminosity vs. log\,$\nu_{peak}$ correlations are studied
simultaneously, a kind of sequence is revealed: at radio frequencies the negative correlation is steep, in optical wavelengths the
correlation is only slightly negative and in the X-ray region it
turns positive. However, the negative correlation of luminosity at 5 GHz and
synchrotron peak frequency is expected to originate solely from the shifting of the
bulk of the synchrotron emission to the higher frequencies with growing
log\,$\nu_{peak}$ and as such does not authenticate the interdependece of source
luminosity and peak frequency. A negative luminosity correlation at 5 GHz and
a positive one at X-rays is what we would expect if the only changing
parameter in the SED sequence was $\nu_{peak}$. When source luminosities at synchrotron peak
frequencies are calculated, we find that they do not correlate with
log\,$\nu_{peak}$. 

In addition to the positive correlation of X-ray luminosity and synchrotron peak
and the lack of synchrotron luminosity correlation,
the blazar sequence scenario is brought into question by the appearance of
numerous low-luminosity LBLs at 5 GHz. The faintest LBLs reach even lower
luminosities than most of the HBLs. As the 5 GHz flux limit of even the least limited
radio-selected sample,
RGB, is 20 mJy, it is obvious that these objects have been identified
for the most part from
X-ray surveys. We note that there are some 5 GHz low--luminosity LBLs that
have a detection at 37 GHz. Especially for these sources we have to consider
the possibility that the 5 GHz flux measurement is from a particularly
quiescent state and does not represent the object accurately. \citet {padovani03} found low-luminosity LBLs among the DXRBS
BL Lacs. In our low-luminosity LBLs there are 4 objects from the DXRBS
survey. 

The 5 GHz low-luminosity LBLs appear also in luminosity plots from other
frequency bands, again, in the low-luminosity end. All the extreme HBLs have relatively high or intermediate
luminosities. This effect, especially in X-rays, seems to support the scenario that UHBLs
are in fact a rare class. If they were common and X-ray luminous, the
numerous X-ray surveys performed lately should have detected them in large
numbers. On the whole, the scatter in luminosities diminishes towards higher
peak frequencies at all wavelengths. There is no solid evidence of truly
high-luminosity HBLs with a measured redshift in radio wavelengths. The high-luminosity HBLs reported by
\citet{giommi05} are from the Sedentary Survey which is not included in our
sample.

We find that the low-luminosity LBLs behave abnormally also on the broad band
spectral index plots, $\alpha_{ro}$\,vs.\,$\alpha_{ox}$, $\alpha_{rx}$\,vs.\,log\,$\nu_{peak}$ and
$\alpha_{ro}$\,vs.\,log\,$\nu_{peak}$. In the first the BL Lac classes take
their expected places except for a few of the low-luminosity LBLs that invade
the region usually populated by HBLs. Doing so, they cross the
$\alpha_{rx}$\,=\,0.75 border frequently used to classify BL Lacs to LBLs and
HBLs. 

When the $\alpha_{rx}$\,vs.\,log\,$\nu_{peak}$ --plot is examined, we find a
strong negative
correlation, but LBLs (log\,$\nu_{peak}<$\,14) disrupt the trend and take on values
$\alpha_{rx}$\,=\,0.45--1. Hints of such a distribution was discovered by
\citet{padovani03}, but the wider range of log\,$\nu_{peak}$ in our sample
makes it evident. All LBLs with conspicuously low $\alpha_{rx}$ are among the
low-luminosity LBLs in the 5 GHz luminosity plot. We considered the
possibility that the low indices in these objects are due to the wider bandwidth of the
ROSAT fluxes compared to the 1 keV fluxes, but this seems to have little
effect. For those objects that had $\alpha_{rx}$ computed for both X-ray
energies, we calculated the average values of both indices. The average for
ROSAT band $\alpha_{rx}$ was 0.63 while for 1 keV fluxes it was 0.69. This
difference is small compared to the deviation of the low--$\alpha_{rx}$ LBLs
from their expected location. Also for 3 of the 10 LBLs with lowest
$\alpha_{rx}$--values 1 keV flux has been used in calculating the spectral index. 

In $\alpha_{ro}$\,vs.\,log\,$\nu_{peak}$ --plot LBLs behave in a similar way,
at log\,$\nu_{peak}\approx$\,13 they have $\alpha_{ro}$=\,0.16--0.95. In this
plot we also see 3 HBLs with conspicuously high $\alpha_{ro}$. They have
relatively high radio luminosities and low optical
luminosities. Each source has only one flux measurement at 5500
$\textrm{\AA}$, possibly from an anomalously quiescent state. For these
HBLs we also considered the possibility that their X-ray flux actually originates
from IC rather than synchrotron radiation. This would lower their peak
frequencies and move them left on the $\alpha_{ro}$\,vs.\,log\,$\nu_{peak}$
--plane. For two of them (RXS J1456.0+5048,
log\,$\nu_{peak}$=19.94 and RXS J1410.5+6100, log\,$\nu_{peak}$=20.25) this
seems possible albeit uncertain.

As explained before, when there are few datapoints representing the SED, the
peak frequency is easily overestimated in the case of extreme HBLs. This
affects also the luminosity correlations. However we do not expect the
significance of the correlations at defined wavelengths to change, even if the
datapoints in the high-energy end moved to the left. The only case
where the consequences could be substantial is the log\,$L_{peak}$
vs. log\,$\nu_{peak}$  --plot (fig.~\ref{peaklum}). If the peak frequency is exaggerated,
so is the luminosity at peak frequency. Thus the uncertain datapoints can move
from upper right to lower left, possibly changing the correlation. We tested
this by artificially lowering the peak luminosities of those sources whose
$\nu F_{peak}$ was notably higher in the SED than their X-ray flux, $\nu
F_{X-ray}$. We changed their peak frequencies accordingly, and the highest
log\,$\nu_{peak}$ in the sample changed from 21.46 to 19.44. Yet the overall
shape of the plot did not change enough to produce significant
correlation. Therefore we find it unlikely that the overestimation of the
log\,$\nu_{peak}$ of HBLs affects the main results of this study.

\section{Conclusions}

In this paper we collected a large amount of multifrequency data as well as new
flux measurements at 37 GHz and plotted the spectral energy distributions of over 300 BL
Lacs of the Mets\"ahovi BL Lac sample. Using such an extensive sample allowed
us to detect the whole range of synchrotron peak frequencies reaching up to
the MeV--region. The main conclusions are as follows:

1. For 22 objects we find that log\,$\nu_{peak}>$\,19. This high--frequency
   tail of the $\nu_{peak}$ distribution is unrepresented in most previous
   studies, but we find it essential for determining the properties of the
   population accurately.

2. The positive correlation of X-ray luminosity and synchrotron peak frequency
   and the lack of correlation between the source luminosity at synchrotron
   peak and peak frequency contradict the blazar sequence scenario. The differences between LBLs and HBLs cannot be attributed to unequal luminosities.

3. When the broad band spectral indices $\alpha_{rx}$ and $\alpha_{ro}$ are
   plotted against log\,$\nu_{peak}$ we
   find substantial scatter in the figures. This implies that the values of
   $\alpha_{rx}$ and $\alpha_{ro}$ cannot be reliably used in BL Lac
   classification.

4. Based on the smooth, declining distribution of $\nu_{peak}$ and the lack of
   any bimodality in all other tests and calculations, we conclude that the BL
   Lac population as a whole is continuous and undivided.


\begin{acknowledgements}
We gratefully acknowledge the funding received from the Academy of Finland for
our Mets\"ahovi and SEST observing projects (project numbers 205969, 46341 and
51436). E.N. thanks Ilona Torniainen for
help in
producing the SED plots for publication.
\end{acknowledgements}  


\bibliographystyle{aa}
\bibliography{enbib}

\Online

\begin{longtable}{l|cccc}
\caption{\label{sample} The Mets\"ahovi BL Lac sample. The synchrotron peak frequency and
  subsequent classification designated in this study are included when
  available.}\\
\hline\hline
Source & RA.(J2000) & Dec.(J2000) & log\,$\nu_{peak}$ & Class \\
\hline\hline
\endfirsthead
\caption{continued.}\\
\hline\hline
Source & RA.(J2000) & Dec.(J2000) & log\,$\nu_{peak}$ & Class \\
\hline\hline
\endhead
NRAO 5 & 00:06:13.9 & -06:23:36 & 12.75 & LBL\\
RX J0007.9+4711 & 00:07:59.9 & 47:12:07 & 16.14 & IBL\\
MS 0011.7+0837 & 00:14:19.7 & 08:54:04 & 16.74 & HBL\\
RXS J0018.4+2947 & 00:18:27.8 & 29:47:32 & - & -\\
PKS 0017+200 & 00:19:37.9 & 20:21:46 & 13.08 & LBL\\
PKS 0019+058 & 00:22:32.5 & 06:08:05 & 13.19 & LBL\\
RXS J0325.2+1515 & 00:35:14.9 & 15:15:04 & 13.73 & LBL\\
1ES 0033+595 & 00:35:52.6 & 59:50:04 & 18.93 & HBL\\
1ES 0037+405 & 00:40:13.8 & 40:50:04 & 16.8 & HBL\\
RXS J0045.3+2127 & 00:45:19.1 & 21:27:43 & 16.89 & HBL\\
B3 0045+395 & 00:47:55.2 & 39:48:57 & - & -\\
EXO 0044.4+2001 & 00:47:08 & 20:17:44 & 16.05 & IBL\\
PKS 0047+023 & 00:49:43.3 & 02:37:04 & 13.56 & LBL\\
PKS 0048-097 & 00:50:41.2 & -09:29:06 & 13.39 & LBL\\
NPM1G -09.0033 & 00:56:20 & -09:36:32 & - & -\\
RXS J0058.2+1723 & 00:58:16.8 & 17:23:14 & - & -\\
Q J0109+181 & 01:09:08.1 & 18:16:03 & - & -\\
NPM1G +41.0022 & 01:10:04.8 & 41:49:50 & 17.7 & HBL\\
RXS J0111.5+0536 & 01:11:30.1 & 05:36:28 & - & -\\
S2 0109+22 & 01:12:05.8 & 22:44:39 & 13.59 & LBL\\
RXS J0115.7+2519 & 01:15:46.5 & 25:19:57 & 13.43 & LBL\\
1ES 0120+340 & 01:23:08.7 & 34:20:51 & 18.32 & HBL\\
MS 0122.1+0903 & 01:24:44.5 & 09:18:49 & 15.53 & IBL\\
B3 0133+388 & 01:36:32.6 & 39:06:00 & 16.59 & HBL\\
PKS 0139-09 & 01:41:25.8 & -09:28:43 & 13.4 & LBL\\
1ES 0145+138 & 01:48:29.7 & 14:02:18 & 15.44 & IBL\\
NPM1G +01.0067 & 01:52:39.6 & 01:47:17 & - & -\\
8C 0149+710 & 01:53:25.8 & 71:15:07 & 14.75 & IBL\\
RXS J0155.9+1502 & 01:56:00.3 & 15:02:13 & - & -\\
87GB 01569+1032 & 01:59:34.4 & 10:47:07 & 15.56 & IBL\\
RXS J0200.4+2712 & 02:00:29.5 & 27:12:36 & - & -\\
MS 0158.5+0019 & 02:01:06.1 & 00:34:00 & 17.87 & HBL\\
RXS J0202.4+0849 & 02:02:26.4 & 08:49:12 & - & -\\
S5 0159+72 & 02:03:33.3 & 72:32:53 & 13.99 & LBL\\
MS 0205.7+3509 & 02:08:38.2 & 35:23:13 & 15.22 & IBL\\
87GB 02109+5130 & 02:14:17.9 & 51:44:52 & 17.69 & HBL\\
RXS J0216.5+2314 & 02:16:32.1 & 23:14:47 & - & -\\
Z 0214+083 & 02:17:17 & 08:37:03 & 15.23 & IBL\\
S4 0218+35 & 02:21:05.4 & 35:56:15 & - & -\\
3C 66A & 02:22:39.6 & 43:02:08 & 15.63 & IBL\\
RXS J0227.2+0201 & 02:27:16.6 & 02:01:58 & - & -\\
1ES 0229+200 & 02:32:48.6 & 20:17:17 & 19.45 & HBL\\
Q 0230+3429 & 02:33:20.3 & 34:42:54 & 15.84 & IBL\\
AO 0235+164 & 02:38:38.8 & 16:36:59 & 13.57 & LBL\\
S5 0238+71 & 02:43:31 & 71:20:18 & 16.3 & IBL\\
NPM1G +10.0097 & 02:45:13.5 & 10:47:23 & - & -\\
RXS J0250.6+1712 & 02:50:38 & 17:12:08 & - & -\\
MS 0257.9+3429 & 03:01:03.8 & 34:41:01 & 13.28 & LBL\\
4C 47.08 & 03:03:35.2 & 47:16:17 & 14.18 & LBL\\
RXS J0303.5+0554 & 03:03:30.1 & 05:54:17 & - & -\\
PKS 0306+102 & 03:09:03.6 & 10:29:16 & 12.94 & LBL\\
VZw331 & 03:13:57.9 & 41:15:24 & 14.48 & LBL\\
RXS J0314.0+2445 & 03:14:02.7 & 24:44:31 & 12.67 & LBL\\
RXS J0314.3+0620 & 03:14:23.9 & 06:19:57 & 19.57 & HBL\\
RXS J0316.1+0904 & 03:16:12.9 & 09:04:43 & 15.91 & IBL\\
MS 03170+1834 & 03:19:51.8 & 18:45:35 & 16.99 & HBL\\
RGB J0321+2336 & 03:22:00 & 23:36:11 & - & -\\
2E 0323+0214 & 03:26:13.9 & 02:25:14 & 19.87 & HBL\\
RXS J0331.3+0654 & 03:31:19.4 & 06:54:28 & - & -\\
RXS J0349.9+0640 & 03:49:59.7 & 06:40:56 & - & -\\
PKS 0406+121 & 04:09:22.1 & 12:17:39 & 13.22 & LBL\\
2E 0414+0057 & 04:16:52.4 & 01:05:24 & 20.71 & HBL\\
1WGA J0421.5+1433 & 04:21:33.1 & 14:33:54 & 13.93 & LBL\\
MS 0419.3+1943 & 04:22:18.5 & 19:50:53 & 16.82 & HBL\\
PKS 0420+022 & 04:22:52.2 & 02:19:27 & - & -\\
PKS 0422+004 & 04:24:46.8 & 00:36:07 & 15.69 & IBL\\
MCG 38364 & 04:25:51.3 & -08:33:38 & 17.17 & HBL\\
2EG J0432+2910 & 04:33:37.7 & 29:05:56 & 14.09 & LBL\\
1ES 0446+449 & 04:50:07.3 & 45:03:12 & - & -\\
PKS 0459+135 & 05:02:33.2 & 13:38:11 & 13.55 & LBL\\
Q 0458+6530 & 05:03:03.4 & 65:34:10 & 18.12 & HBL\\
RXS J0505.5+0416 & 05:05:34.7 & 04:15:54 & 16.94 & HBL\\
1ES 0502+675 & 05:07:56.1 & 67:37:24 & 19.18 & HBL\\
S5 0454+84 & 05:08:42.5 & 84:32:05 & 13.58 & LBL\\
MG 0509+0541 & 05:09:25.9 & 05:41:35 & 15.34 & IBL\\
4U 0506-03 & 05:09:39 & -04:00:36 & 17.94 & HBL\\
2E 0514+0626 & 05:17:04 & 06:29:39 & - & -\\
1ES 0525+713 & 05:31:41.7 & 71:22:17 & - & -\\
TEX 0554+534 & 05:58:11.6 & 53:28:19 & 14.44 & LBL\\
MS 0607.9+7108 & 06:13:42.8 & 71:07:29 & 14.85 & IBL\\
87GB 06216+4441 & 06:25:18.3 & 44:40:02 & 13.61 & LBL\\
1ES 0647+250 & 06:50:46.5 & 25:03:00 & 18.28 & HBL\\
B3 0651+428 & 06:54:43.5 & 42:47:59 & 15.12 & IBL\\
NPM1G +42.0131 & 06:56:10.6 & 42:37:02 & 17.25 & HBL\\
EXO 0706.1+5913 & 07:10:30.1 & 59:08:21 & 21.05 & HBL\\
RXS J0712.3+5719 & 07:12:18.7 & 57:19:22 & - & -\\
S5 0716+714 & 07:21:53.3 & 71:20:36 & 14.46 & LBL\\
RXS J0723.2+5841 & 07:23:13.2 & 58:41:23 & - & -\\
FIRST J0724.7+2621 & 07:24:42.8 & 26:21:30 & 16.39 & IBL\\
PKS 0723-008 & 07:25:50.7 & -00:54:56 & - & -\\
FIRST J0730.4+3307 & 07:30:26.1 & 33:07:22 & 16.29 & IBL\\
RXS J0737.3+3517 & 07:37:21 & 35:17:41 & 17.77 & HBL\\
FIRST J0738.6+3139 & 07:38:37.8 & 31:39:30 & - & -\\
PKS 0735+17 & 07:38:07.4 & 17:42:19 & 13.95 & LBL\\
FIRST J0741.3+2253 & 07:41:18.8 & 22:53:39 & - & -\\
MS 0737.9+7441 & 07:44:05.1 & 74:33:59 & 13.61 & LBL\\
S4 0749+54 & 07:53:01.3 & 53:53:00 & 13.12 & LBL\\
GB 0751+485 & 07:54:45.7 & 48:23:51 & 14.32 & LBL\\
PKS 0754+100 & 07:57:06.7 & 09:56:35 & 13.63 & LBL\\
RXS J0800.1+6210 & 08:00:06.5 & 62:10:12 & - & -\\
RXS J0801.0+6444 & 08:01:00.7 & 64:44:43 & - & -\\
RXS J0805.4+7534 & 08:05:26.5 & 75:34:25 & 15.96 & IBL\\
SBS 0802+596 & 08:06:25.9 & 59:31:07 & 16.69 & HBL\\
B2 0806+31 & 08:09:13.4 & 31:22:22 & - & -\\
RXS J0809.6+3455 & 08:09:38.5 & 34:55:37 & 18.29 & HBL\\
1ES 0806+524 & 08:09:49.2 & 52:18:58 & 16.56 & HBL\\
PKS 0808+019 & 08:11:26.6 & 01:46:52 & 13.17 & LBL\\
RXS J0812.1+5717 & 08:12:08.8 & 57:17:34 & - & -\\
EXO 0811.2+2949 & 08:14:21.8 & 29:40:32 & - & -\\
1WGA J0816.0-0736 & 08:16:04.3 & -07:35:57 & 14.19 & LBL\\
RXS J0816.3+5739 & 08:16:23.8 & 57:39:03 & 17.19 & HBL\\
RXS J0816.6+6208 & 08:16:40.9 & 62:08:44 & 15.25 & IBL\\
OJ 425 & 08:18:16.1 & 42:22:46 & 13.33 & LBL\\
FIRST J0818.4+2814 & 08:18:27.3 & 28:14:02 & 16.01 & IBL\\
FIRST J0819.4+4037 & 08:19:25.8 & 40:37:43 & 16.61 & HBL\\
FIRST J0820.3+3640 & 08:20:20.2 & 36:40:04 & 14 & LBL\\
4C 22.21 & 08:23:24.8 & 22:23:03 & 13.09 & LBL\\
PKS 0823+033 & 08:25:50.3 & 03:09:24 & 13.08 & LBL\\
RXS J0828.2+4153 & 08:28:14.2 & 41:53:50 & 18.66 & HBL\\
B3 0827+395 & 08:30:19.4 & 39:23:47 & - & -\\
PKS 0829+046 & 08:31:48.9 & 04:29:39 & 13.53 & LBL\\
1H 0827+089 & 08:31:54.8 & 08:47:58 & 14.22 & LBL\\
OJ 448 & 08:32:23.2 & 49:13:21 & 13.06 & LBL\\
TEX 0836+182 & 08:39:30.7 & 18:02:47 & 14.55 & IBL\\
FIRST J0847.0+4117 & 08:47:02.5 & 41:17:57 & 18.11 & HBL\\
RXS J0847.2+1133 & 08:47:12.9 & 11:33:52 & 19.13 & HBL\\
RXS J0848.4+8111 & 08:48:27.8 & 81:11:47 & - & -\\
US 1889 & 08:54:09.9 & 44:08:31 & 17.42 & HBL\\
OJ 287 & 08:54:48.8 & 20:06:30 & 13.89 & LBL\\
NPM1G -09.0307 & 09:08:02.2 & -09:59:37 & 15.52 & IBL\\
B2 0906+31 & 09:09:53.3 & 31:06:03 & 17.4 & HBL\\
Ton 1015 & 09:10:37.1 & 33:29:24 & 15.39 & IBL\\
FIRST J0910.8+3902 & 09:10:52 & 39:02:02 & 18.33 & HBL\\
B2 0912+29 & 09:15:52.4 & 29:33:24 & 16 & IBL\\
RXS J0916.8+5238 & 09:16:52 & 52:38:28 & 17.22 & HBL\\
MS 0922.9+7459 & 09:28:02.6 & 74:47:19 & - & -\\
RXS J0929.2+5013 & 09:29:15.4 & 50:13:35 & 14.59 & IBL\\
S5 0916+86 & 09:29:42.7 & 86:12:21 & 14.16 & LBL\\
1ES 0927+500 & 09:30:37.5 & 49:50:25 & 21.13 & HBL\\
B2 0927+35 & 09:30:55.3 & 35:03:38 & 14.8 & IBL\\
RXS J0930.9+3933 & 09:30:56.8 & 39:33:33 & - & -\\
SBS 0936+522 & 09:39:37.9 & 52:01:46 & - & -\\
B2 0937+26 & 09:40:13.6 & 26:03:26 & 14.75 & IBL\\
US 1015 & 09:50:11.8 & 45:53:20 & 15.48 & IBL\\
RGB J0952+656 & 09:52:32.2 & 65:38:01 & 15.08 & IBL\\
MS 0950.9+4929 & 09:54:09.8 & 49:15:00 & 16.92 & HBL\\
S4 0954+65 & 09:58:47.2 & 65:33:54 & 13.76 & LBL\\
4C 22.25 & 10:00:21.36 & 22:33:07.4 & - & -\\
MS 0958.9+2102 & 10:01:42.4 & 20:48:18 & 15.54 & IBL\\
EXO 1004.0+3509 & 10:06:56.3 & 34:54:45 & 16.92 & HBL\\
RXS J10081+4705 & 10:08:11.3 & 47:05:20 & 19.67 & HBL\\
NRAO 350 & 10:12:13.3 & 06:30:57 & 16.09 & IBL\\
RXS J1012.7+4229 & 10:12:44.3 & 42:29:57 & 20.97 & HBL\\
GB 1011+496 & 10:15:04.2 & 49:26:01 & 16.74 & HBL\\
RXS J1016.2+4108 & 10:16:16.7 & 41:08:13 & 16.62 & HBL\\
RXS J1022.7-0112 & 10:22:43.9 & -01:12:56 & 17.97 & HBL\\
1ES 1028+511 & 10:31:18.5 & 50:53:36 & 18.56 & HBL\\
FIRST J1032.6+3738 & 10:32:40.7 & 37:38:26 & - & -\\
RXS J1037.7+5711 & 10:37:44.2 & 57:11:57 & 14.95 & IBL\\
TEX 1040+244 & 10:43:09 & 24:08:35 & 13.1 & LBL\\
1ES 1044+549 & 10:47:45.8 & 54:37:41 & 13 & LBL\\
MS 1050.7+4946 & 10:53:44.2 & 49:29:54 & 15.29 & IBL\\
FIRST J1054.5+3855 & 10:54:31.8 & 38:55:22 & 16.68 & HBL\\
RXS J1055.5-0126 & 10:55:34.1 & -01:26:05 & - & -\\
RXS J1056.1+0252 & 10:56:06.6 & 02:52:13 & - & -\\
FIRST J1057.3+2303 & 10:57:23.1 & 23:03:18 & 18.7 & HBL\\
RXS J1057.8+0059 & 10:57:52.4 & 00:59:13 & - & -\\
FIRST J1058.4+2817 & 10:58:29.9 & 28:17:46 & 18.37 & HBL\\
RXS J1058.6+5628 & 10:58:37.7 & 56:28:12 & 15.64 & IBL\\
MC 1057+100 & 11:00:20.2 & 09:49:35 & - & -\\
RXS J1100.3+4019 & 11:00:21.1 & 40:19:28 & 18.76 & HBL\\
RXS J1102.8-0148 & 11:02:52 & -01:48:51 & - & -\\
MRK 421 & 11:04:27.2 & 38:12:32 & 18.49 & HBL\\
1ES 1106+244 & 11:09:16.2 & 24:11:20 & 16.91 & HBL\\
OP 1106.7+3654 & 11:09:33.5 & 36:38:26 & - & -\\
FIRST J1110.9+3539 & 11:10:56.9 & 35:39:06 & - & -\\
RXS J1110.6+7133 & 11:10:37.5 & 71:33:57 & 16.96 & HBL\\
RXS J1111.5+3452 & 11:11:30.9 & 34:52:00 & - & -\\
FIRST J1117.6+2548 & 11:17:40.4 & 25:48:46 & 15.71 & IBL\\
EXO 1118.0+4228 & 11:20:48.1 & 42:12:13 & 17.41 & HBL\\
US 2504 & 11:29:50.1 & 26:52:53 & 12.97 & LBL\\
MS 1133.7+1618 & 11:36:17.6 & 16:01:53 & 15.9 & IBL\\
MRK 180 & 11:36:26.5 & 70:09:28 & 18.61 & HBL\\
RXS J1136.5+6737 & 11:36:30.1 & 67:37:04 & 17.55 & HBL\\
FIRST J1136.8+2550 & 11:36:50.1 & 25:50:52 & 15.13 & IBL\\
2E 1146+2456 & 11:49:29.9 & 24:38:55 & 17.87 & HBL\\
EXO 1449.9+2455 & 11:49:30.3 & 24:39:27 & 19.83 & HBL\\
B2 1147+245 & 11:50:19.2 & 24:17:54 & 13.95 & LBL\\
RXS J1151.4+5859 & 11:51:24.6 & 58:59:14 & 16.4 & IBL\\
FIRST J1152.1+2837 & 11:52:10.7 & 28:37:21 & 14.8 & IBL\\
FIRST J1153.7+3823 & 11:53:42.9 & 38:23:06 & - & -\\
MS 1154.1+4255 & 11:56:46.6 & 42:38:10 & 14.91 & IBL\\
B3 1159+450 & 12:02:08.6 & 44:44:21 & 15.92 & IBL\\
B3 1206+416 & 12:09:22.8 & 41:19:41 & 14.59 & IBL\\
1207+39W4 & 12:10:26.6 & 39:29:08 & - & -\\
MS 1209.0+3917 & 12:11:34.2 & 39:00:55 & - & -\\
1ES 1212+078 & 12:15:10.9 & 07:32:03 & 15.91 & IBL\\
Q 1214+1753 & 12:16:56.9 & 17:37:12 & - & -\\
B2 1215+30 & 12:17:52 & 30:07:01 & 15.58 & IBL\\
GB2 1217+348 & 12:20:08.4 & 34:31:22 & 14.46 & LBL\\
PG 1218+304 & 12:21:21.9 & 30:10:37 & 19.14 & HBL\\
ON 231 & 12:21:31.7 & 28:13:58 & 14.84 & IBL\\
UM 493 & 12:22:06.5 & -01:06:38 & - & -\\
RXS J1222.2+3541 & 12:22:12.4 & 35:41:00 & 12.81 & LBL\\
S5 1221+80 & 12:23:40.4 & 80:40:04 & 14.21 & LBL\\
MS 1221.8+2452 & 12:24:24.3 & 24:36:24 & 13.99 & LBL\\
1WGA J1225.3+1818 & 12:25:18.2 & 18:18:20 & 14.71 & IBL\\
FIRST J1226.0+2604 & 12:26:04.1 & 26:04:28 & 16.29 & IBL\\
RXS J1230.2+2517 & 12:30:14 & 25:18:06 & 14.9 & IBL\\
2E 1258+1437 & 12:31:23.9 & 14:21:25 & 14.91 & IBL\\
MS 1229.2+6430 & 12:31:31.3 & 64:14:18 & 16.25 & IBL\\
B2 1229+29 & 12:31:43.6 & 28:47:49 & - & -\\
FIRST J1236.3+3900 & 12:36:23.1 & 39:00:01 & 16.61 & HBL\\
MS 1235.4+6315 & 12:37:38.6 & 62:58:44 & 15.98 & IBL\\
RXS J1237.0+3020 & 12:37:05.5 & 30:20:04 & - & -\\
1ES 1239+069 & 12:41:48.3 & 06:36:01 & 17.38 & HBL\\
RXS J1241.6+3440 & 12:41:41.2 & 34:40:32 & - & -\\
Ton 116 & 12:43:12.7 & 36:27:44 & - & -\\
PG 1246+586 & 12:48:18.8 & 58:20:29 & 15.05 & IBL\\
1ES 1249+174E & 12:51:45.5 & 17:11:17 & - & -\\
FIRST J1252.3+2640 & 12:52:19.5 & 26:40:53 & - & -\\
S4 1250+53 & 12:53:11.9 & 53:01:11 & 14.82 & IBL\\
1ES 1255+244 & 12:57:31.9 & 24:12:40 & 16.89 & HBL\\
MS 1256.3+0151 & 12:58:54.6 & 01:34:43 & - & -\\
MS 1258.4+6401 & 13:00:17.6 & 63:44:39 & 16.35 & IBL\\
FIRST J1301.7+4056 & 13:01:45.7 & 40:56:24 & 16.55 & HBL\\
RXS J1302.9+5056 & 13:02:55.5 & 50:56:17 & - & -\\
MC2 1307+12 & 13:09:33.9 & 11:54:24 & 13.07 & LBL\\
1WGA J1309.6+0828 & 13:09:38.9 & 08:28:28 & 14.64 & IBL\\
AUCVn & 13:10:28.66 & 32:30:43.8 & 13.75 & LBL\\
HS 1309+2605 & 13:12:19.2 & 25:49:58 & - & -\\
TEX 1312+240 & 13:14:43.8 & 23:48:26 & 15.84 & IBL\\
RXS J1319.5+1405 & 13:19:31.7 & 14:05:34 & 20.85 & HBL\\
UM 566 & 13:19:55.1 & 01:52:58 & - & -\\
1ES 1320+084N & 13:22:54.9 & 08:10:10 & 13.12 & LBL\\
RXS J1324.0+5739 & 13:24:00.8 & 57:39:16 & 15.48 & IBL\\
RXS J1326.2+2933 & 13:26:15 & 29:33:29 & - & -\\
RXS J1326.2+1230 & 13:26:17.6 & 12:30:00 & 16.32 & IBL\\
RX J1340.1+2743 & 13:40:10.9 & 27:43:48 & - & -\\
RXS J1340.4+4410 & 13:40:29.8 & 44:10:04 & 16.51 & HBL\\
RXS J1341.0+3959 & 13:41:05 & 39:59:45 & 20.06 & HBL\\
1338.8+2705 & 13:41:05.8 & 26:50:26 & - & -\\
1340.8+2721 & 13:43:05.1 & 27:06:24 & - & -\\
RXS J1353.4+5601 & 13:53:28 & 56:00:55 & 19.23 & HBL\\
FIRST J1354.4+3706 & 13:54:26.7 & 37:06:54 & 16.92 & HBL\\
RXS J1359.8+5911 & 13:59:53.7 & 59:11:01 & 13.66 & LBL\\
MC 1400+162 & 14:02:44.5 & 15:59:57 & 16.47 & IBL\\
RXS J1404.8+6554 & 14:04:49.6 & 65:54:30 & - & -\\
MS 1402.3+0416 & 14:04:51 & 04:02:02 & 15.83 & IBL\\
MS 1407.9+5954 & 14:09:23.5 & 59:39:41 & 16.63 & HBL\\
PKS 1407+022 & 14:10:04.6 & 02:03:07 & 13.69 & LBL\\
RXS J1410.5+6100 & 14:10:31.7 & 61:00:10 & 20.25 & HBL\\
FIRST J1414.1+3430 & 14:14:09.3 & 34:30:57 & - & -\\
RGB J1415+485 & 14:15:36.8 & 48:30:30 & 14.56 & IBL\\
PKS 1413+135 & 14:15:58.8 & 13:20:24 & 12.83 & LBL\\
CRSS 1416.3+1137 & 14:16:20.7 & 11:37:37 & - & -\\
2E 1415+2557 & 14:17:56.6 & 25:43:25 & 19.24 & HBL\\
OQ 530 & 14:19:46.6 & 54:23:14 & 14.16 & LBL\\
RXS J1422.6+5801 & 14:22:39 & 58:01:55 & - & -\\
FIRST J1426.1+3404 & 14:26:07.7 & 34:04:26 & 14.1 & LBL\\
PKS 1424+240 & 14:27:00.5 & 23:48:00 & 15.7 & IBL\\
RGB J1427+541 & 14:27:30.3 & 54:09:24 & 14.89 & IBL\\
H 1426+428 & 14:28:32.7 & 42:40:21 & 18.55 & HBL\\
TEX 1428+370 & 14:30:40.6 & 36:49:03 & 14.26 & LBL\\
CSO 474 & 14:36:45.7 & 35:57:01 & - & -\\
RXS J1436.9+5639 & 14:36:57.8 & 56:39:25 & 17.5 & HBL\\
PG 1437+398 & 14:39:17.5 & 39:32:43 & 16.7 & HBL\\
1ES 1440+122 & 14:42:48.3 & 12:00:40 & 16.45 & IBL\\
MS 1443.5+6349 & 14:44:34.9 & 63:36:06 & 17.05 & HBL\\
RXS J1445.0-0326 & 14:45:06.2 & -03:26:12 & - & -\\
RXS J1448.0+3608 & 14:48:00.6 & 36:08:31 & 16.73 & HBL\\
RXS J1449.5+2746 & 14:49:32.8 & 27:46:21 & 18.82 & HBL\\
RXS J1451 4+6354 & 14:51:27.5 & 63:54:19 & - & -\\
SBS 1452+516 & 14:54:27.1 & 51:24:33 & - & -\\
RXS J1456.0+5048 & 14:56:03.7 & 50:48:25 & 19.94 & HBL\\
RXS J1458.4+4832 & 14:58:28 & 48:32:40 & 21.46 & HBL\\
B3 1456+375 & 14:58:44.8 & 37:20:22 & 13.47 & LBL\\
MS 1458.8+2249 & 15:01:01.9 & 22:38:06 & 15.26 & IBL\\
FIRST J1502.1+2528 & 15:02:08.3 & 25:28:45 & 14.17 & LBL\\
FIRST J1502.5+3350 & 15:02:34 & 33:50:55 & 13.66 & LBL\\
RXS J1508.7+2709 & 15:08:42.9 & 27:09:10 & 17.3 & HBL\\
SBS 1508+561 & 15:09:48 & 55:56:17 & 15.22 & IBL\\
FIRST J1515.9+2426 & 15:15:56.2 & 24:26:20 & 15.89 & IBL\\
RXS J1516.7+2918 & 15:16:41.6 & 29:18:10 & 18.62 & HBL\\
PKS 1514+197 & 15:16:56.8 & 19:32:12 & 13.6 & LBL\\
1H 1515+660 & 15:17:47.6 & 65:25:24 & 18.11 & HBL\\
FIRST J1530.7+2310 & 15:30:44 & 23:10:13 & - & -\\
FAQS J1530.7+5329 & 15:30:44.5 & 53:29:28 & 16.88 & HBL\\
RXS J1532.0+3016 & 15:32:02.2 & 30:16:29 & 17.05 & HBL\\
RXS J1533.4+3416 & 15:33:24.3 & 34:16:41 & 18.32 & HBL\\
RGB J1534+372 & 15:34:47.2 & 37:15:55 & 14.26 & LBL\\
1ES 1533+535 & 15:35:00.8 & 53:20:37 & 19.68 & HBL\\
FIRST J1535.4+3922 & 15:35:29.1 & 39:22:46 & 16.38 & IBL\\
MS 1534.2+0148 & 15:36:46.8 & 01:37:59 & 18.83 & HBL\\
1ES 1544+820 & 15:40:15.8 & 81:55:07 & 17.79 & HBL\\
4C 14.6 & 15:40:46.5 & 14:47:45.9 & 14.85 & IBL\\
RXS J1542.9+6129 & 15:42:56.9 & 61:29:56 & 14.72 & IBL\\
RXS J1544.3+0458 & 15:44:18.7 & 04:58:24 & 16.77 & HBL\\
MS 1552.1+2020 & 15:54:24.1 & 20:11:25 & 17.12 & HBL\\
PG 1553+11 & 15:55:43.1 & 11:11:24 & 16.49 & IBL\\
MYC 1557+566 & 15:58:48.5 & 56:25:14 & - & -\\
RXS J1602.2+3050 & 16:02:18 & 30:51:09 & 16.42 & IBL\\
PKS 1604+159 & 16:07:06.4 & 15:51:34 & 14.73 & IBL\\
RXS J1610.0+6710 & 16:10:04.1 & 67:10:26 & 17.45 & HBL\\
B3 1615+412 & 16:17:06.6 & 41:06:45 & 14.41 & LBL\\
87GB 16166+2206 & 16:18:47.9 & 21:59:26 & 15.66 & IBL\\
4C 37.46 & 16:21:11.3 & 37:46:04.9 & - & -\\
RXJ 1624.7+3726 & 16:24:43.4 & 37:26:42 & - & -\\
RXS J1624.9+7554 & 16:24:56.5 & 75:54:55 & 13.3 & LBL\\
RXS J1626.4+3513 & 16:26:25.6 & 35:13:38 & 15.28 & IBL\\
RXS J1631.3+4217 & 16:31:24.7 & 42:17:03 & 18.99 & HBL\\
RXS J1638.0+7326 & 16:38:02.6 & 73:26:10 & - & -\\
RXS J1644.2+4546 & 16:44:20 & 45:46:44 & 17.48 & HBL\\
1643.2+4021 & 16:44:53.2 & 40:16:28 & - & -\\
FIRST J1645.9+2947 & 16:45:57.7 & 29:47:30 & - & -\\
FIRST J1651.1+4212 & 16:51:09.2 & 42:12:53 & 17.82 & HBL\\
RXS J1651.6+7218 & 16:51:41.5 & 72:18:19 & - & -\\
RGB J1652+403 & 16:52:50 & 40:23:10 & 14.97 & IBL\\
MRK 501 & 16:53:52.2 & 39:45:36 & 16.84 & HBL\\
B3 1659+399 & 17:01:24.6 & 39:54:36 & 13.41 & LBL\\
FIRST J1702.1+2643 & 17:02:09.6 & 26:43:15 & 14.27 & LBL\\
FIRST J1702.6+3115 & 17:02:38.6 & 31:15:43 & 13.49 & LBL\\
RXS J1704.8+7138 & 17:04:46.9 & 71:38:18 & 15.58 & IBL\\
MS 1704.9+6046 & 17:05:34.9 & 60:42:17 & 17.17 & HBL\\
FIRST J1712.8+2931 & 17:12:48.8 & 29:31:17 & 18.79 & HBL\\
RXS J1718.6+7358 & 17:18:40.1 & 73:58:15 & - & -\\
PKS 1717+177 & 17:19:13.1 & 17:45:07 & 13.08 & LBL\\
RXS J1719.6+7443 & 17:19:41 & 74:43:00 & 14.19 & LBL\\
B2 1722+40 & 17:24:05.5 & 40:04:38 & 13 & LBL\\
H 1722+119 & 17:25:04.4 & 11:52:16 & 15.8 & IBL\\
IZw187 & 17:28:18.6 & 50:13:11 & 17.42 & HBL\\
RXS J1728.6+7041 & 17:28:38.3 & 70:41:08 & 13.6 & LBL\\
RXS J1732.0+6926 & 17:32:05.4 & 69:26:16 & - & -\\
OT 465 & 17:39:57.1 & 47:37:59 & 13.91 & LBL\\
RGBJ 1742+597 & 17:42:32 & 59:45:06 & 14.18 & LBL\\
NPM1G +19.0510 & 17:43:57.9 & 19:35:09 & 17.91 & HBL\\
B3 1743+398B & 17:45:37.7 & 39:51:31 & 17.71 & HBL\\
1ES 1745+504 & 17:46:32.3 & 50:28:09 & - & -\\
B3 1746+470 & 17:47:26.6 & 46:58:51 & 13.96 & LBL\\
S4 1749+70 & 17:48:33.1 & 70:05:50 & 14.55 & IBL\\
B3 1747+433 & 17:49:00.4 & 43:21:52 & 13.56 & LBL\\
RXS J1750.0+4700 & 17:50:05 & 47:00:44 & 18.1 & HBL\\
PKS 1749+096 & 17:51:32.7 & 09:39:01 & 12.78 & LBL\\
RXS J1756.2+5522 & 17:56:15.9 & 55:22:18 & 19.9 & HBL\\
MS 1757.7+7034 & 17:57:13.3 & 70:33:37 & 13.7 & LBL\\
S5 1803+784 & 18:00:45.4 & 78:28:04 & 14.05 & LBL\\
RX J1801.7+6638 & 18:01:46.7 & 66:38:40 & - & -\\
3C 371 & 18:06:50.7 & 69:49:28 & 14.65 & IBL\\
RGB J1808+468 & 18:08:01.2 & 46:49:41 & 14.63 & IBL\\
RGB J1811+442 & 18:11:53.5 & 44:16:29 & 15.07 & IBL\\
B2 1811+31 & 18:13:35.3 & 31:44:17 & 15.53 & IBL\\
4C 56.27 & 18:24:07.07 & 56:51:01.5 & 12.95 & LBL\\
RXS J1829.4+5402 & 18:29:24.3 & 54:03:00 & 15.07 & IBL\\
Q 1832+6845 & 18:32:36.6 & 68:48:09 & - & -\\
RXS J1838.7+4802 & 18:38:49.2 & 48:02:34 & 13.52 & LBL\\
RGB J1841+591 & 18:41:20.3 & 59:06:08 & 14.91 & IBL\\
RXS J1848.7+4245 & 18:48:47.1 & 42:45:39 & - & -\\
1ES 1853+671 & 18:53:52.1 & 67:13:55 & 16.64 & HBL\\
87GB 19021+5536 & 19:03:11.6 & 55:40:39 & 14.5 & LBL\\
S4 1926+61 & 19:27:30.4 & 61:17:32 & 13.44 & LBL\\
RXS J1931.1+0937 & 19:31:09.5 & 09:37:13 & - & -\\
1ES 1959+650 & 19:59:59.9 & 65:08:55 & 18.03 & HBL\\
S5 2007+77 & 20:05:31.1 & 77:52:43 & 13.15 & LBL\\
S5 2010+72 & 20:09:52.3 & 72:29:19 & 13.64 & LBL\\
PKS 2012-017 & 20:15:15.1 & -01:37:33 & 14.68 & IBL\\
S5 2023+76 & 20:22:35.6 & 76:11:26 & 14.1 & LBL\\
PKS 2032+107 & 20:35:22.3 & 10:56:06 & 14 & LBL\\
1ES 2037+521 & 20:39:23.5 & 52:19:50 & 16.13 & IBL\\
PKS 2047+039 & 20:50:06.2 & 04:07:49 & 13.83 & LBL\\
S5 2051+74 & 20:51:33.8 & 74:41:40 & 18.75 & HBL\\
PKS 2131-021 & 21:34:10.2 & -01:53:17 & 12.76 & LBL\\
MS 2143.4+0704 & 21:45:52.3 & 07:19:27 & 13.92 & LBL\\
PKS 2149+17 & 21:52:24.8 & 17:34:39 & 13.85 & LBL\\
BL LAC & 22:02:43.3 & 42:16:39 & 14.28 & LBL\\
RXS J2209.3+1031 & 22:09:18.5 & 10:31:43 & 13.36 & LBL\\
RXS J2219.7+2120 & 22:19:45.3 & 21:20:48 & 17.61 & HBL\\
RXS J2225.1+1136 & 22:25:11.2 & 11:36:01 & - & -\\
PKS 2223-114 & 22:25:43.7 & -11:13:41 & 14.65 & IBL\\
3C 446 & 22:25:45.1 & -04:56:34 & 13.37 & LBL\\
S5 2229+69 & 22:30:35.6 & 69:46:29 & 13.09 & LBL\\
RXS J2233.0+1335 & 22:33:00.9 & 13:35:59 & 16.61 & HBL\\
RGB J2243+203 & 22:43:54.7 & 20:21:04 & 14.15 & LBL\\
B3 2247+381 & 22:50:05.8 & 38:24:37 & 15.61 & IBL\\
PKS 2254+074 & 22:57:17.3 & 07:43:12 & 14.18 & LBL\\
RXS J2304.6+3705 & 23:04:36.6 & 37:05:08 & 21.01 & HBL\\
B3 2311+396A & 23:13:50.2 & 40:03:03 & - & -\\
Q J2319+161 & 23:19:43.4 & 16:11:51 & 15.48 & IBL\\
TEX 2320+343 & 23:22:44 & 34:36:14 & 16.73 & HBL\\
1ES 2321+419 & 23:23:54.1 & 42:11:19 & 13.26 & LBL\\
B3 2322+396 & 23:25:17.9 & 39:57:37 & 15.29 & IBL\\
1ES 2326+174 & 23:29:03.3 & 17:43:30 & 18.07 & HBL\\
Q J2338+212 & 23:38:56.4 & 21:24:41 & 17.62 & HBL\\
MS 2336.5+0517 & 23:39:07 & 05:34:36 & 14.91 & IBL\\
1ES 2344+514 & 23:47:04.8 & 51:42:18 & 16.4 & IBL\\
MS 2347.4+1924 & 23:50:01.7 & 19:41:52 & 15.8 & IBL\\
RXS J2350.3-059 & 23:50:17.9 & -05:59:28 & - & -\\
TEX 2348+360 & 23:50:36.7 & 36:22:11 & 16.1 & IBL\\
PKS 2354-02 & 23:57:25.1 & -01:52:15 & - & -\\
\hline\hline
\end{longtable}

\begin{figure*}
\centering
\includegraphics[width=17cm]{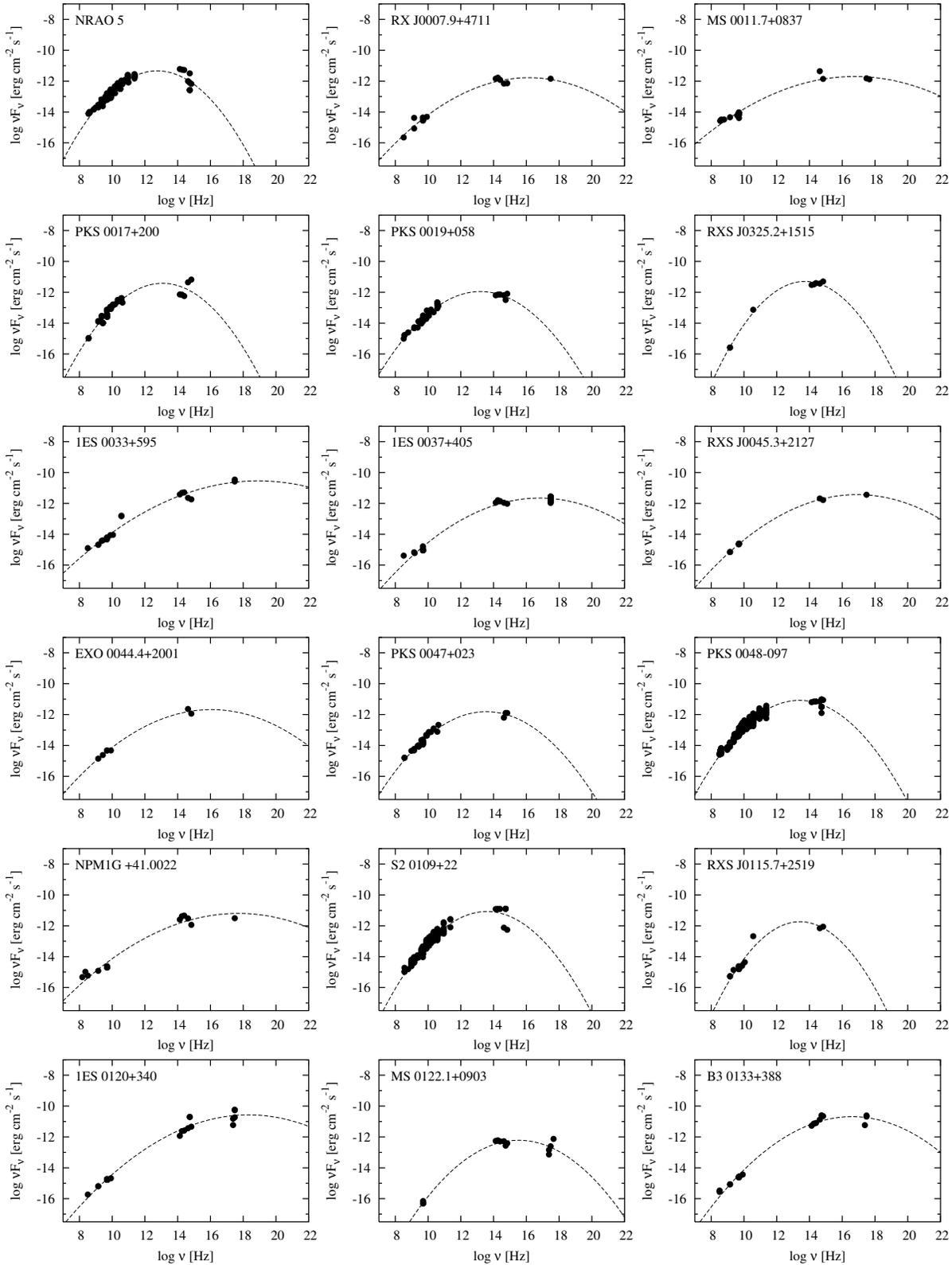}
\caption{Spectral energy distributions of the Mets\"ahovi BL Lac sample. Only
  datapoints used in the fit are shown in the figure.}
\label{sed1}
\end{figure*}

\addtocounter{figure}{-1}
\begin{figure*}
\centering
\includegraphics[width=17cm]{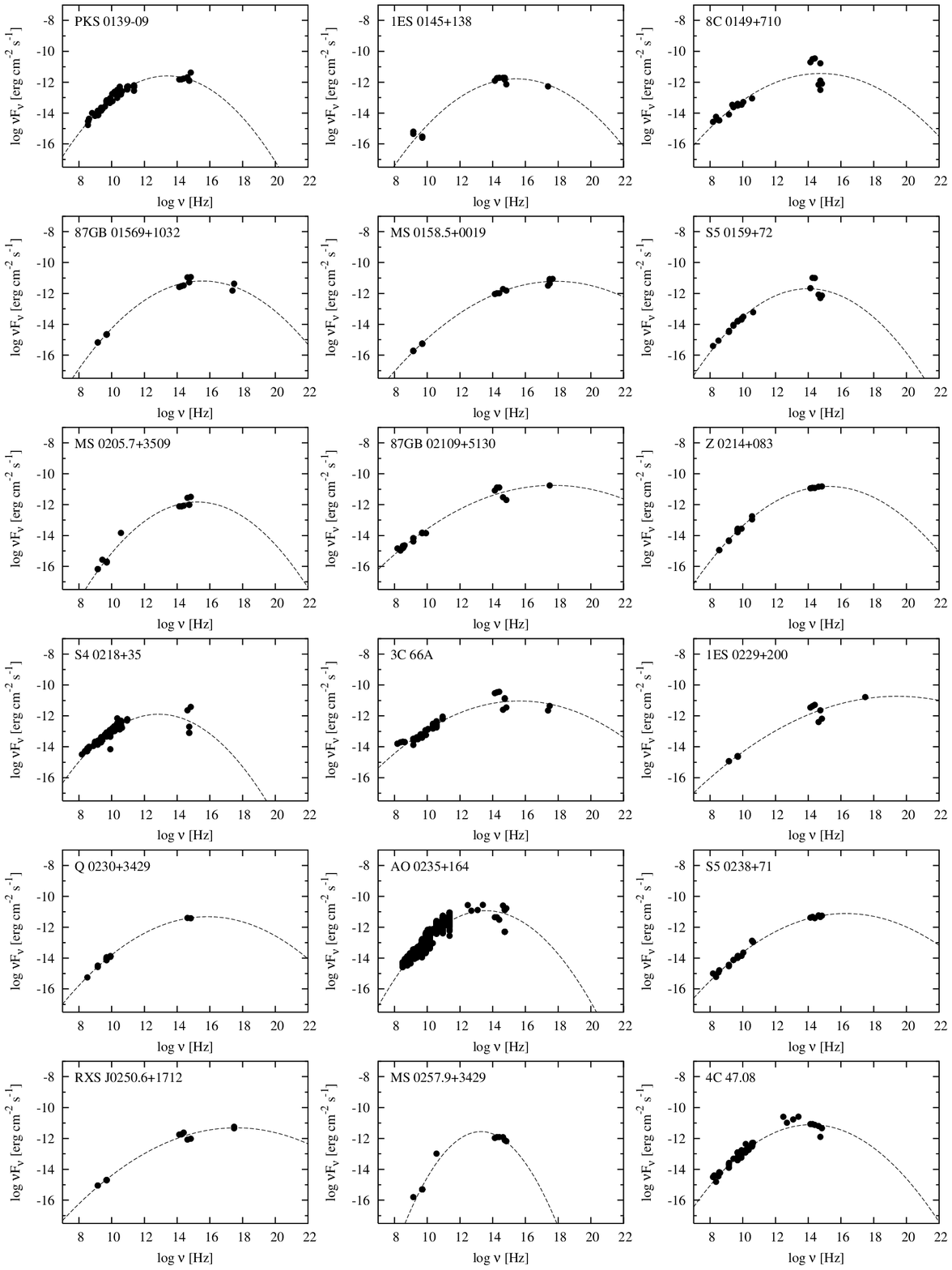}
\caption{continued.}
\label{sed2}
\end{figure*}

\addtocounter{figure}{-1}
\begin{figure*}
\centering
\includegraphics[width=17cm]{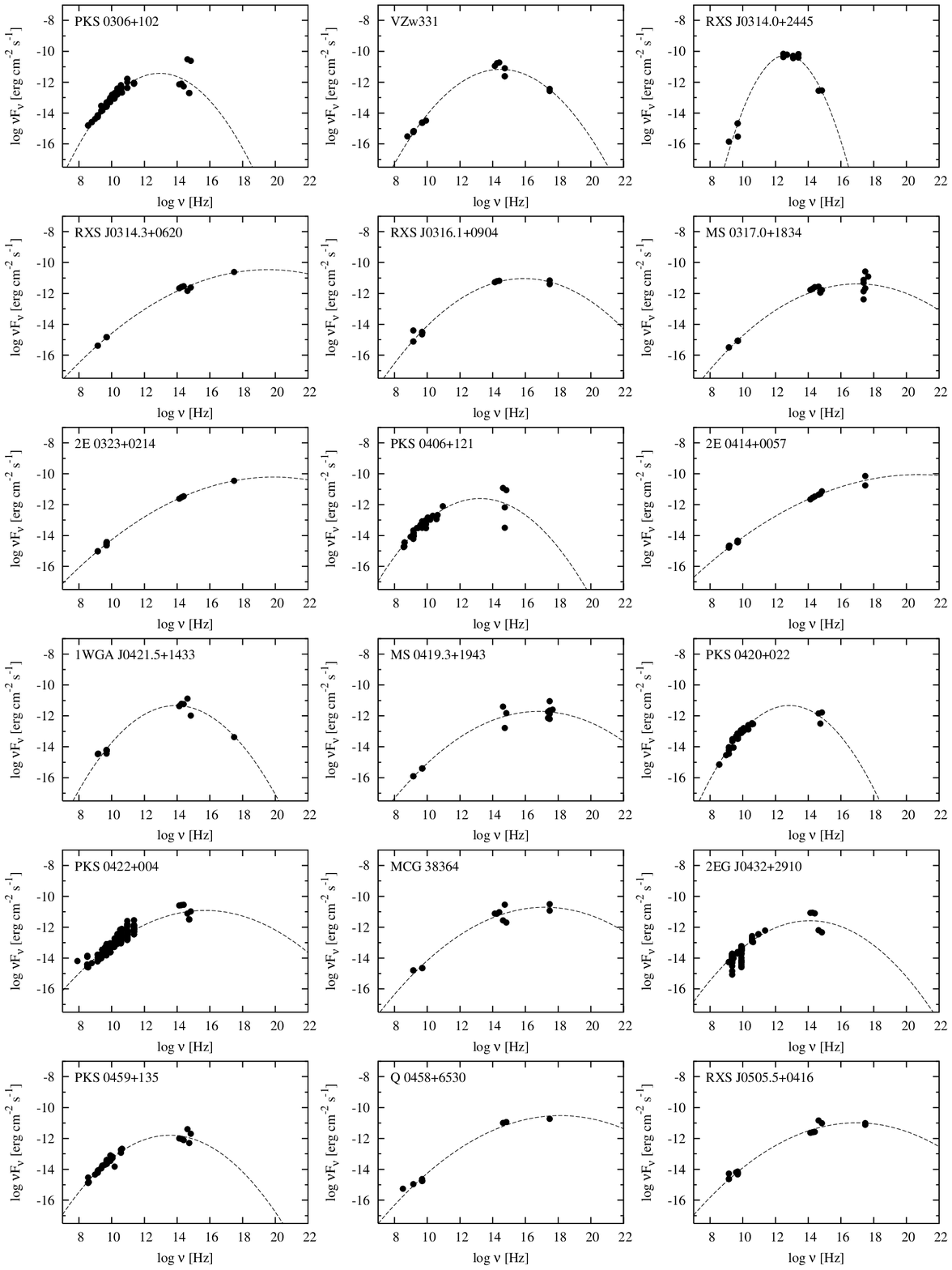}
\caption{continued.}
\label{sed3}
\end{figure*}

\addtocounter{figure}{-1}
\begin{figure*}
\centering
\includegraphics[width=17cm]{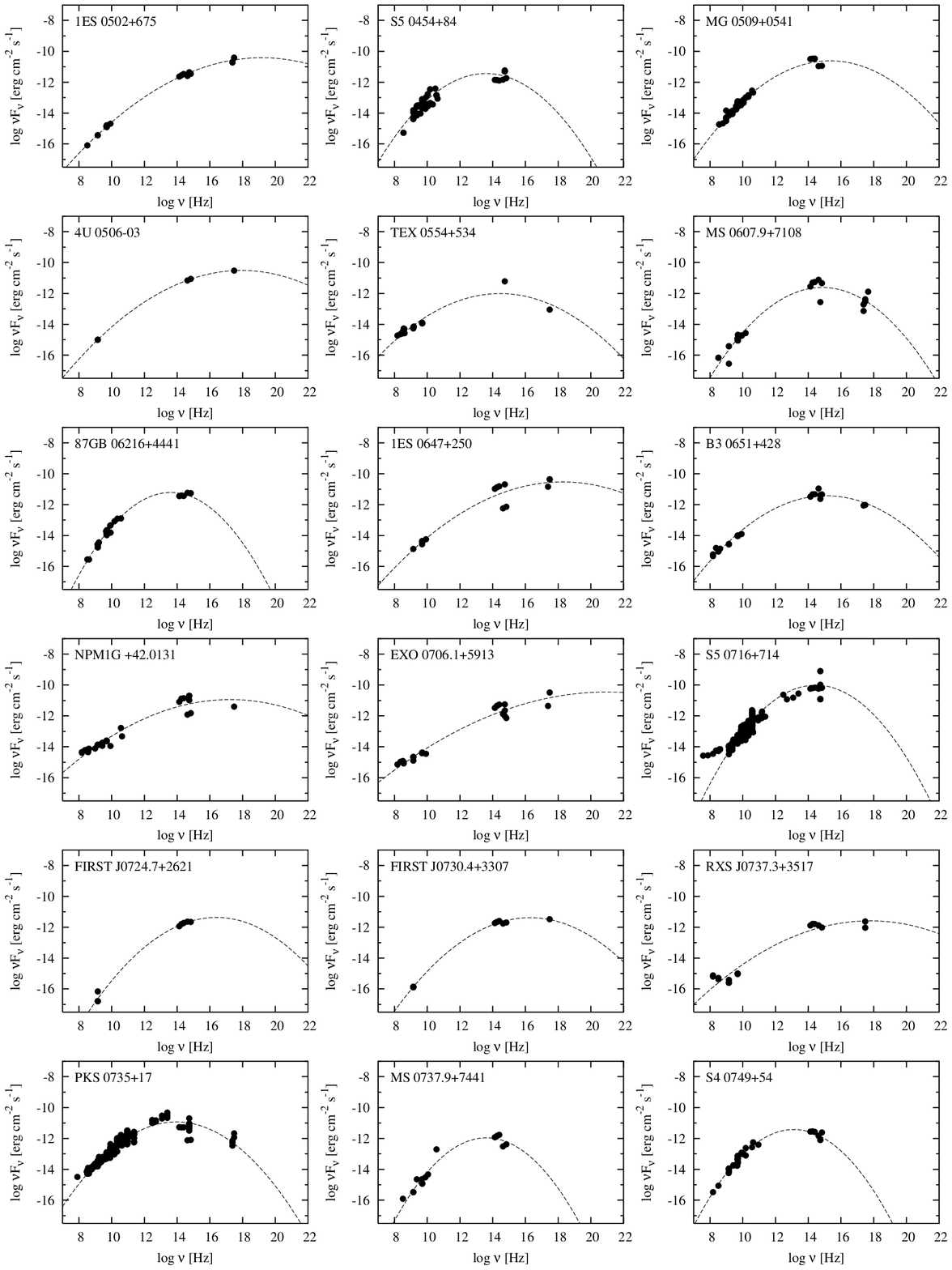}
\caption{continued.}
\label{sed4}
\end{figure*}

\addtocounter{figure}{-1}
\begin{figure*}
\centering
\includegraphics[width=17cm]{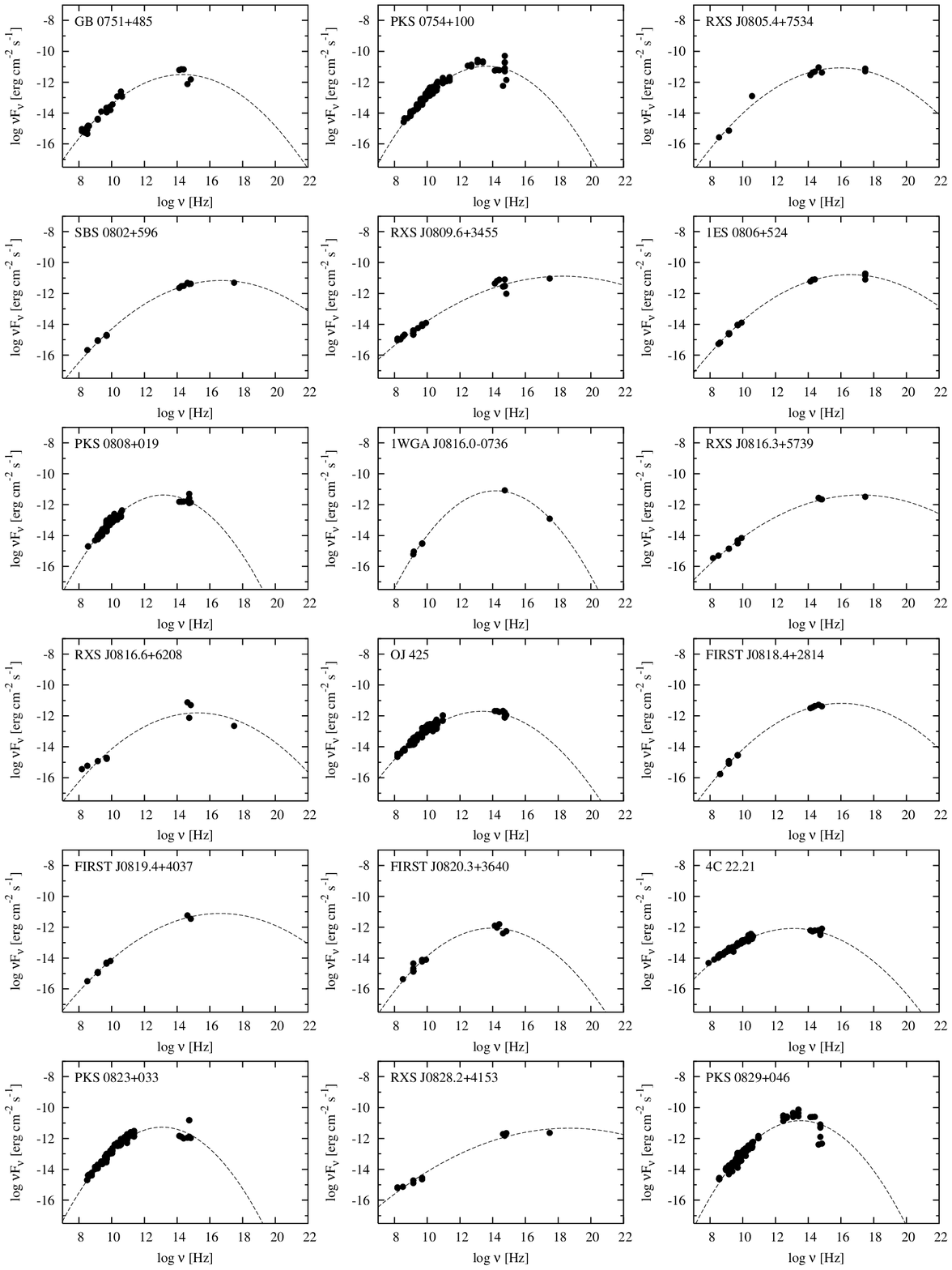}
\caption{continued.}
\label{sed5}
\end{figure*}

\addtocounter{figure}{-1}
\begin{figure*}
\centering
\includegraphics[width=17cm]{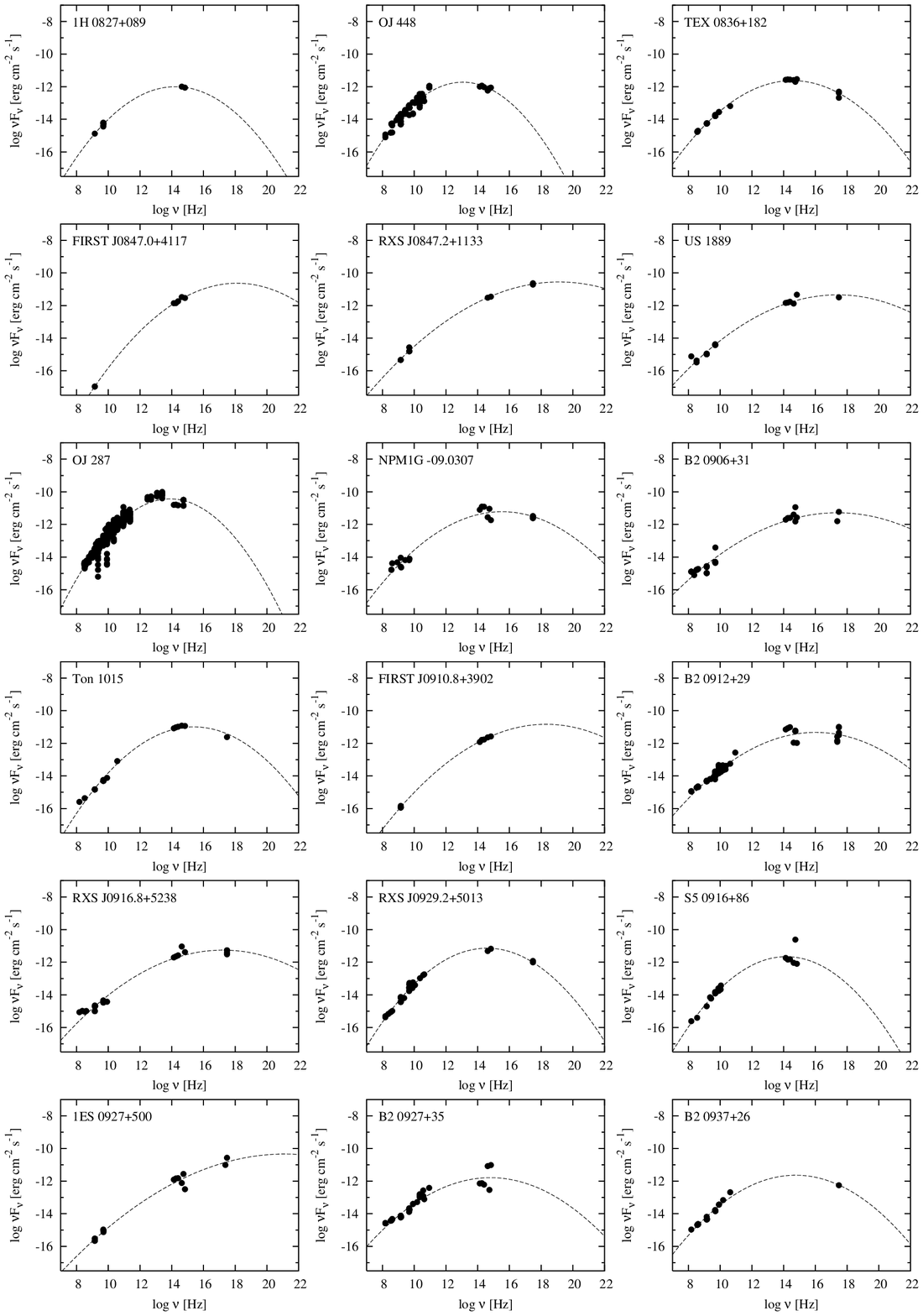}
\caption{continued.}
\label{sed6}
\end{figure*}

\addtocounter{figure}{-1}
\begin{figure*}
\centering
\includegraphics[width=17cm]{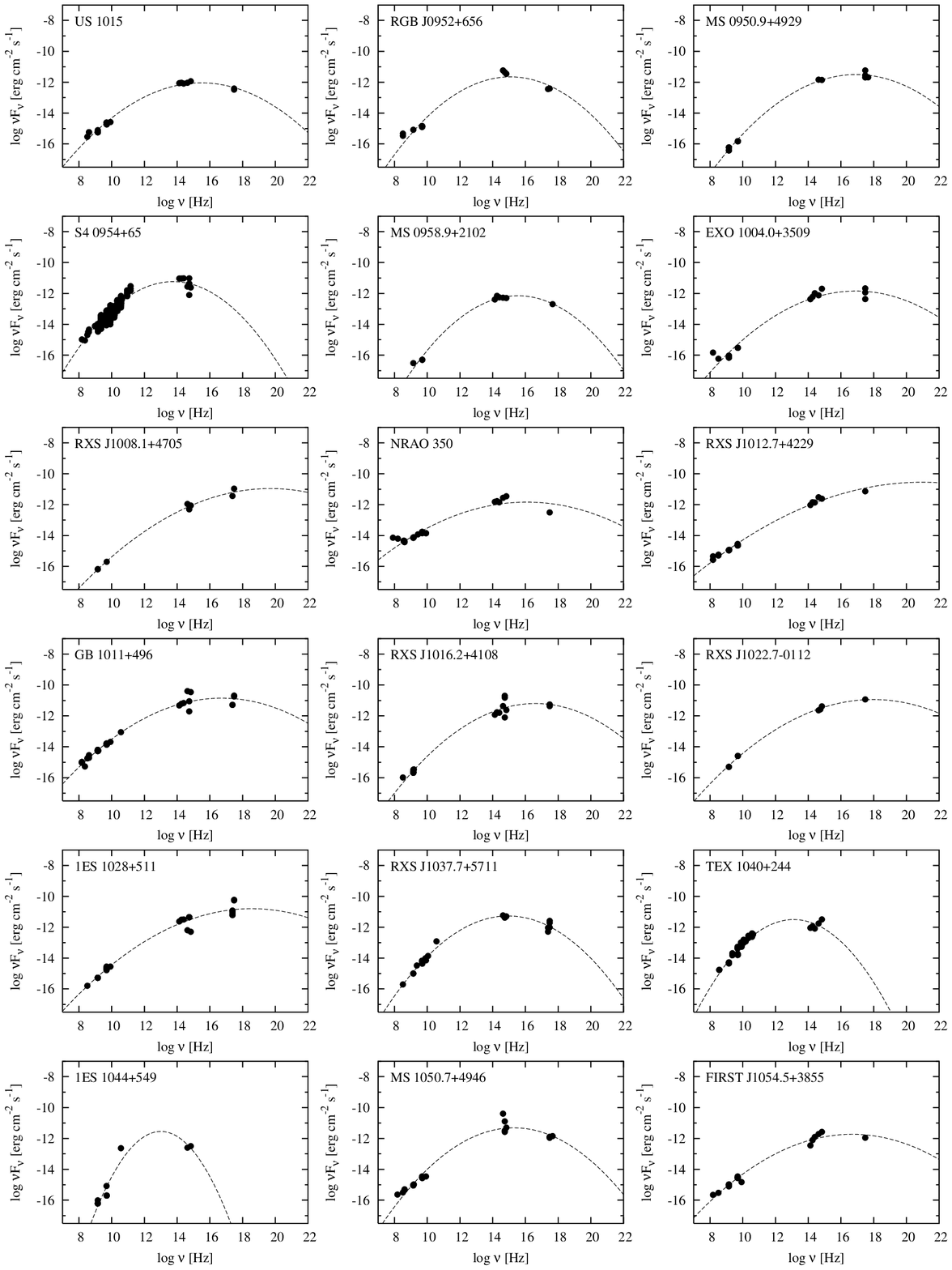}
\caption{continued.}
\label{sed7}
\end{figure*}

\addtocounter{figure}{-1}
\begin{figure*}
\centering
\includegraphics[width=17cm]{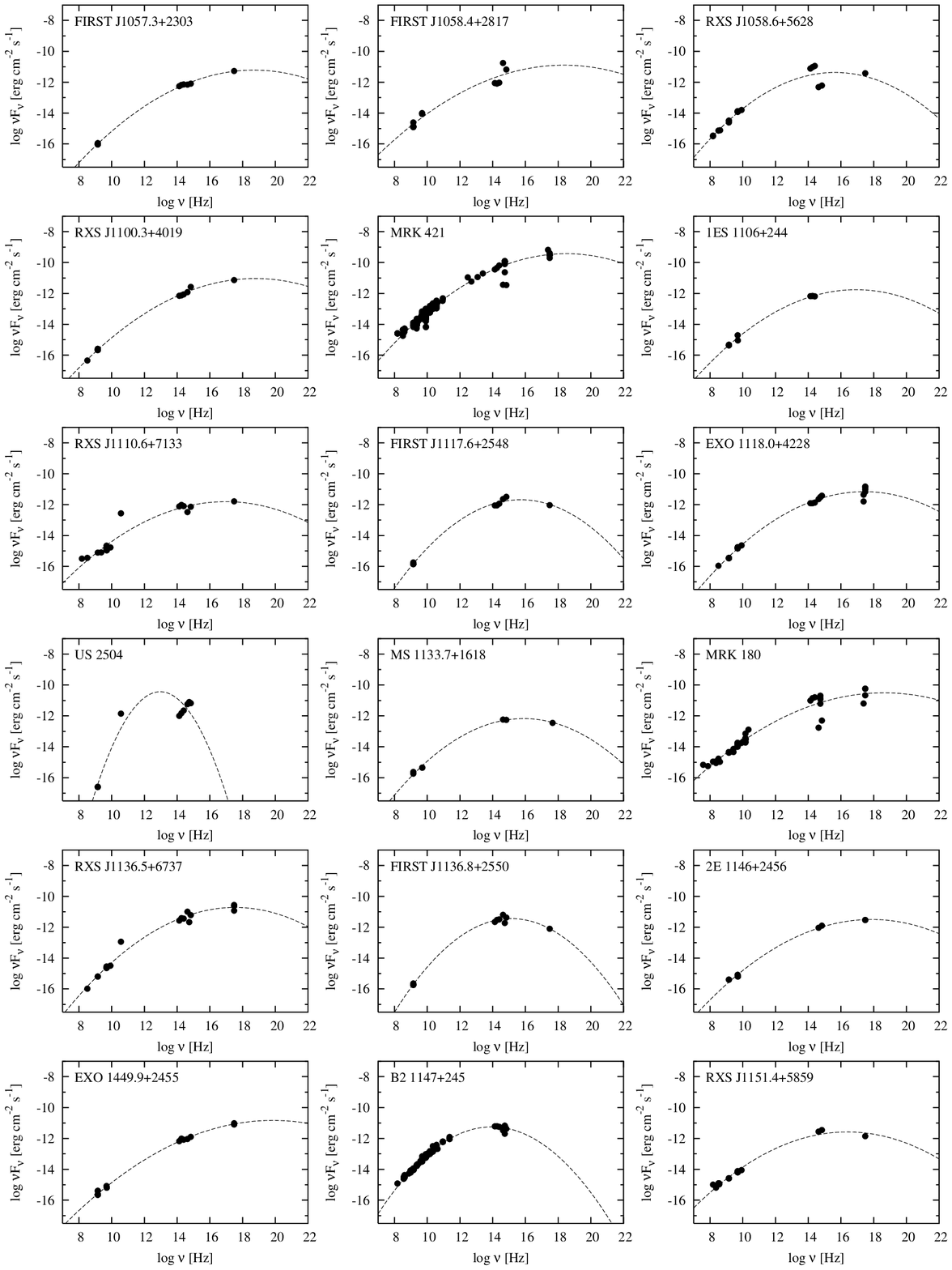}
\caption{continued.}
\label{sed8}
\end{figure*}

\addtocounter{figure}{-1}
\begin{figure*}
\centering
\includegraphics[width=17cm]{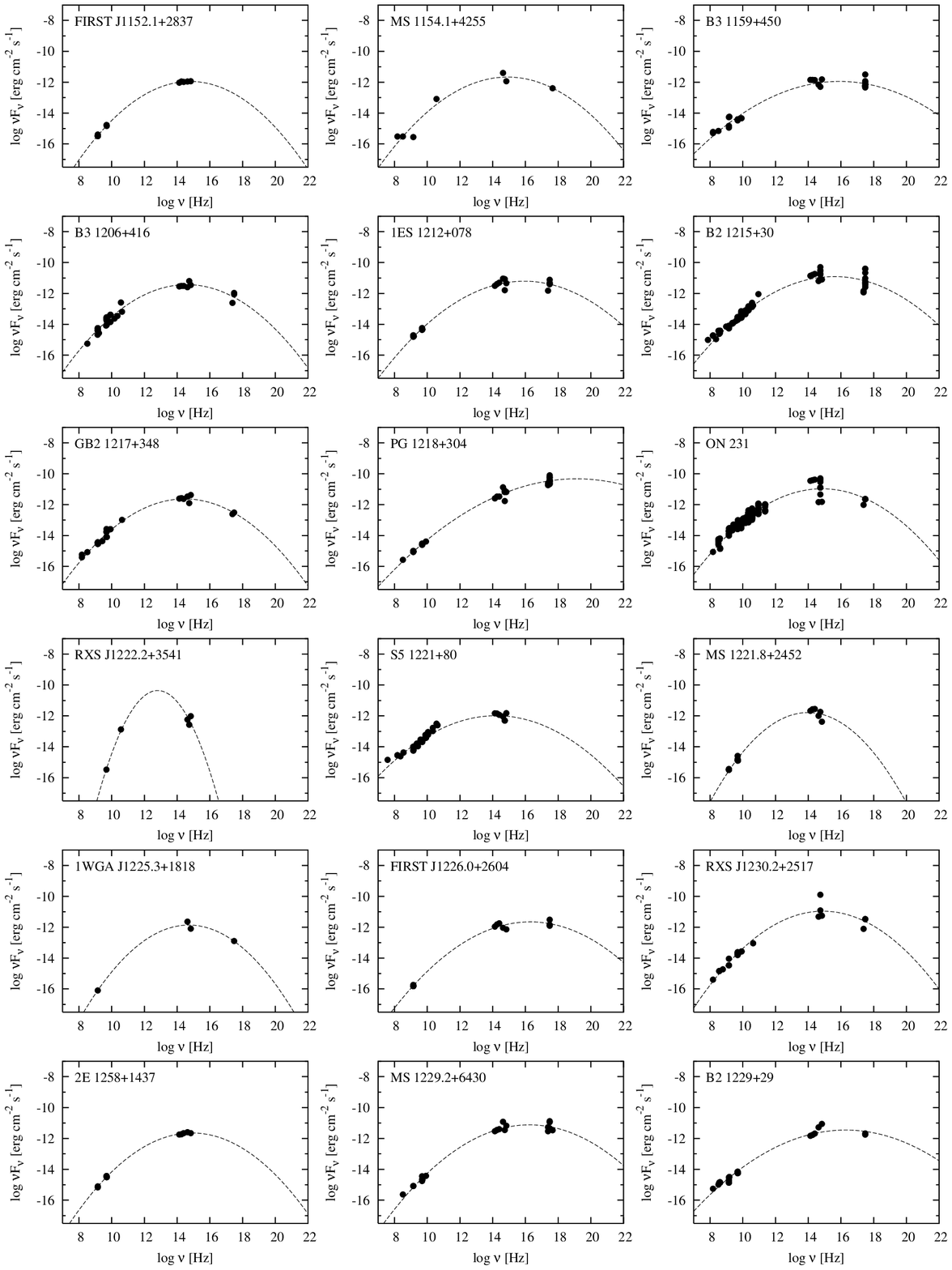}
\caption{continued.}
\label{sed9}
\end{figure*}

\addtocounter{figure}{-1}
\begin{figure*}
\centering
\includegraphics[width=17cm]{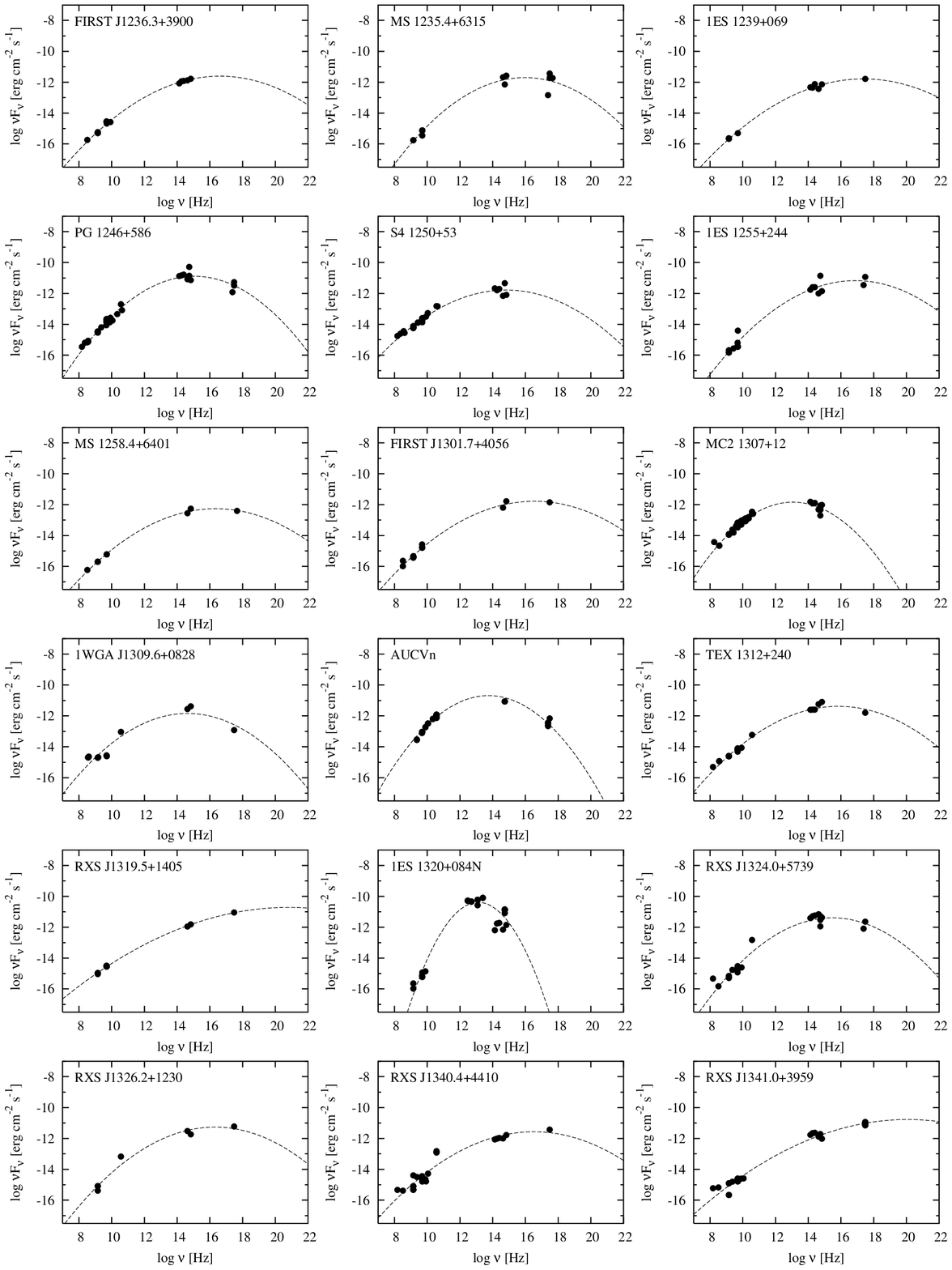}
\caption{continued.}
\label{sed10}
\end{figure*}

\addtocounter{figure}{-1}
\begin{figure*}
\centering
\includegraphics[width=17cm]{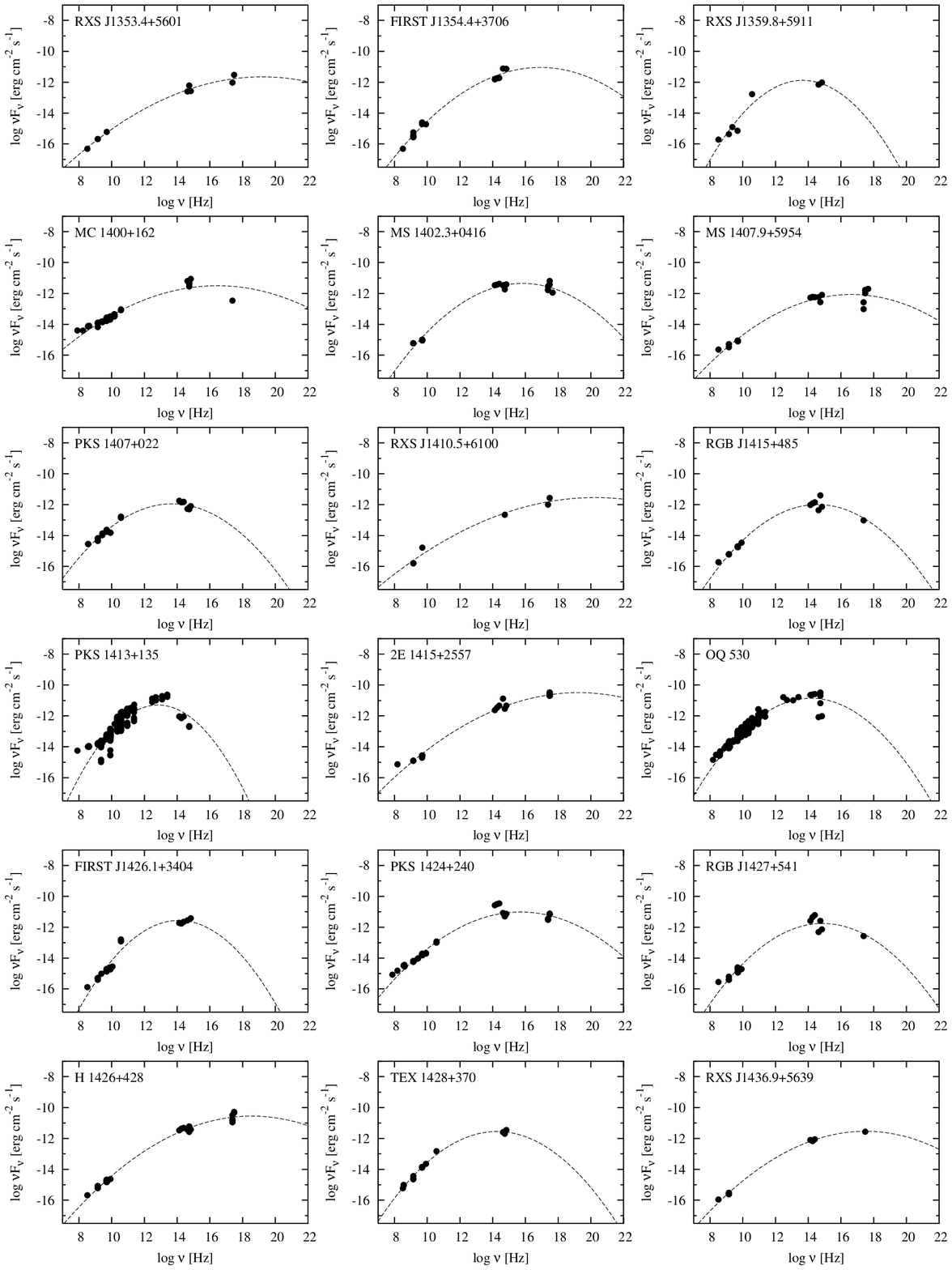}
\caption{continued.}
\label{sed11}
\end{figure*}

\addtocounter{figure}{-1}
\begin{figure*}
\centering
\includegraphics[width=17cm]{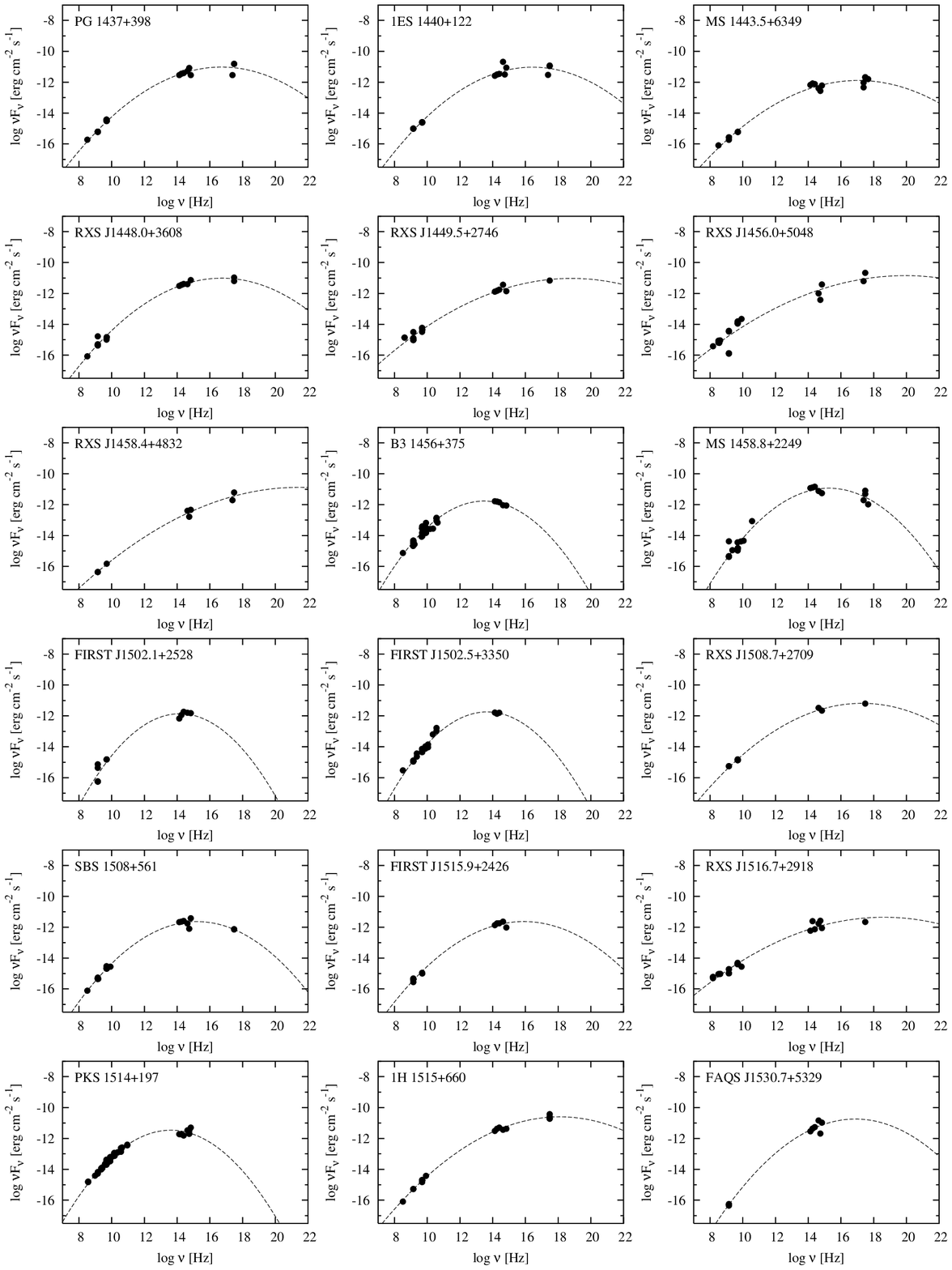}
\caption{continued.}
\label{sed12}
\end{figure*}

\addtocounter{figure}{-1}
\begin{figure*}
\centering
\includegraphics[width=17cm]{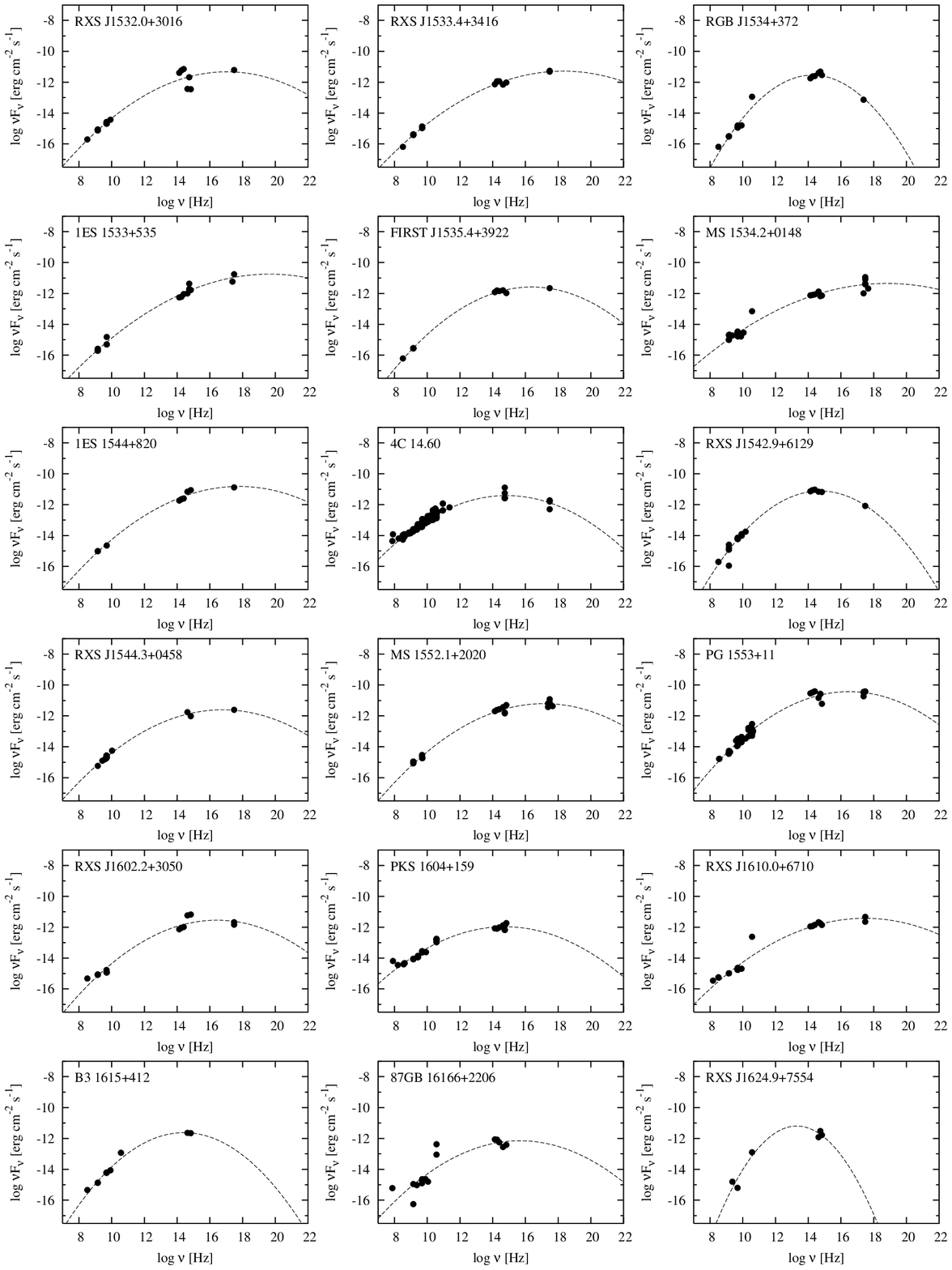}
\caption{continued.}
\label{sed13}
\end{figure*}

\addtocounter{figure}{-1}
\begin{figure*}
\centering
\includegraphics[width=17cm]{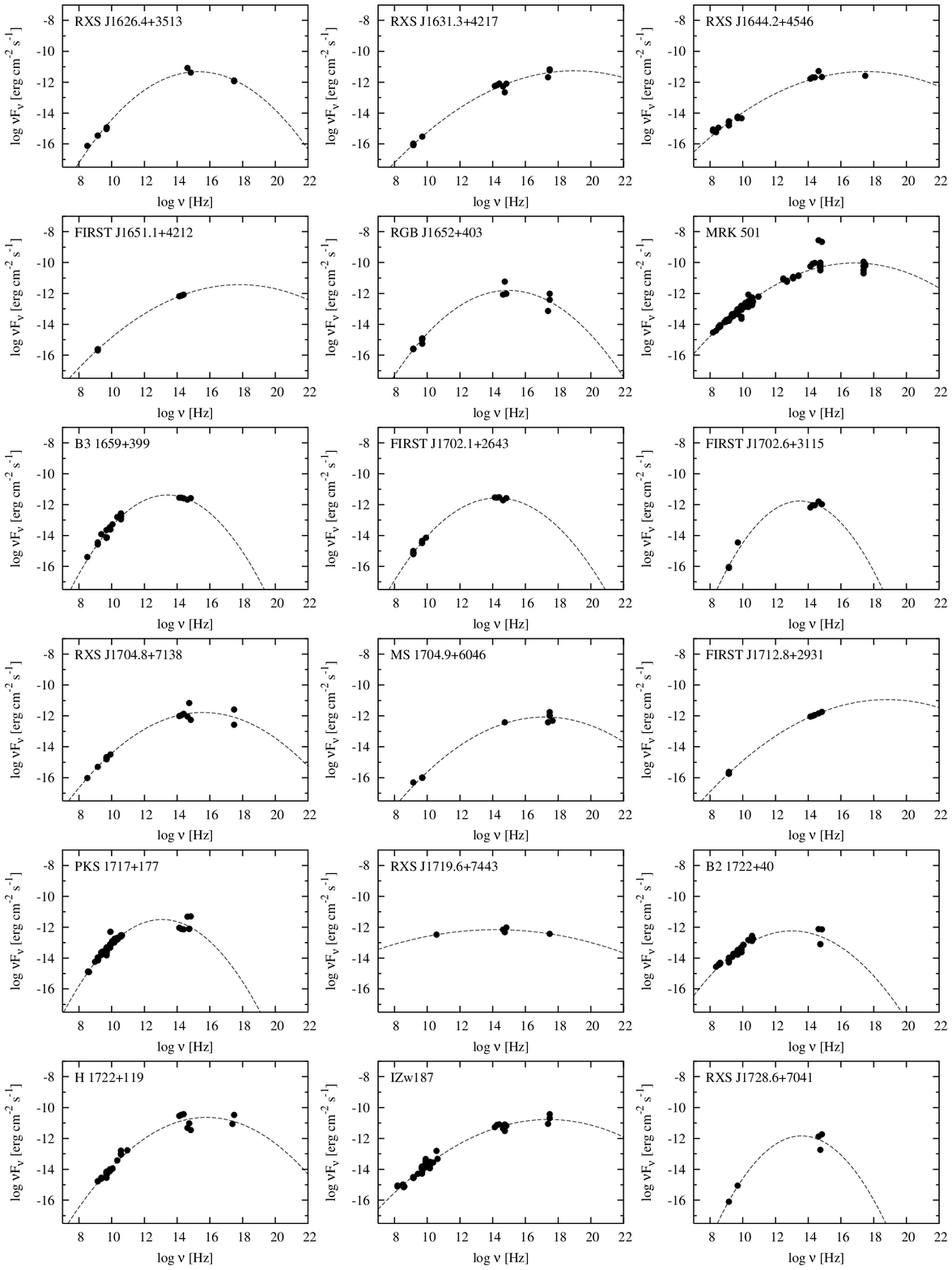}
\caption{continued.}
\label{sed14}
\end{figure*}

\addtocounter{figure}{-1}
\begin{figure*}
\centering
\includegraphics[width=17cm]{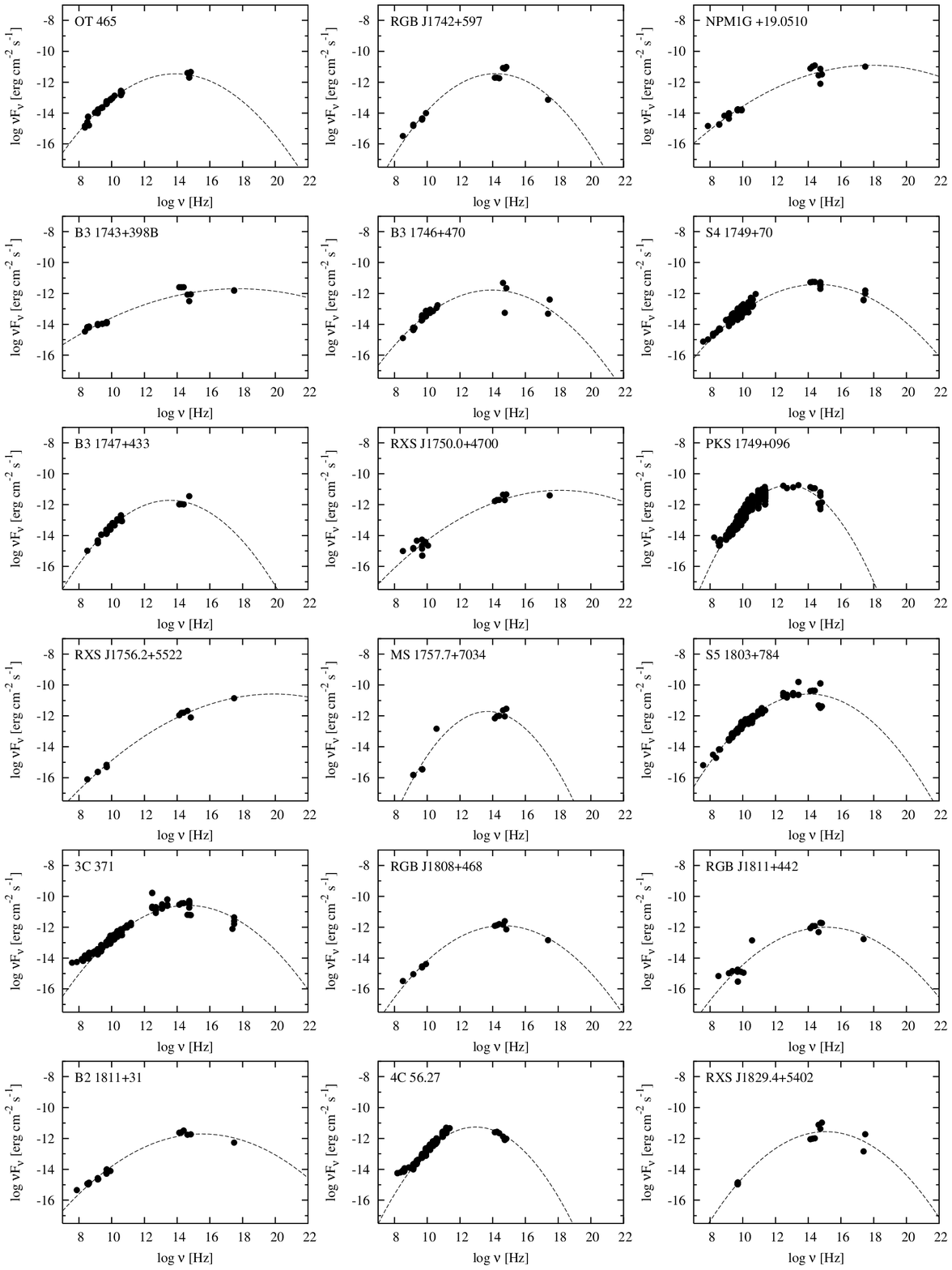}
\caption{continued.}
\label{sed15}
\end{figure*}

\addtocounter{figure}{-1}
\begin{figure*}
\centering
\includegraphics[width=17cm]{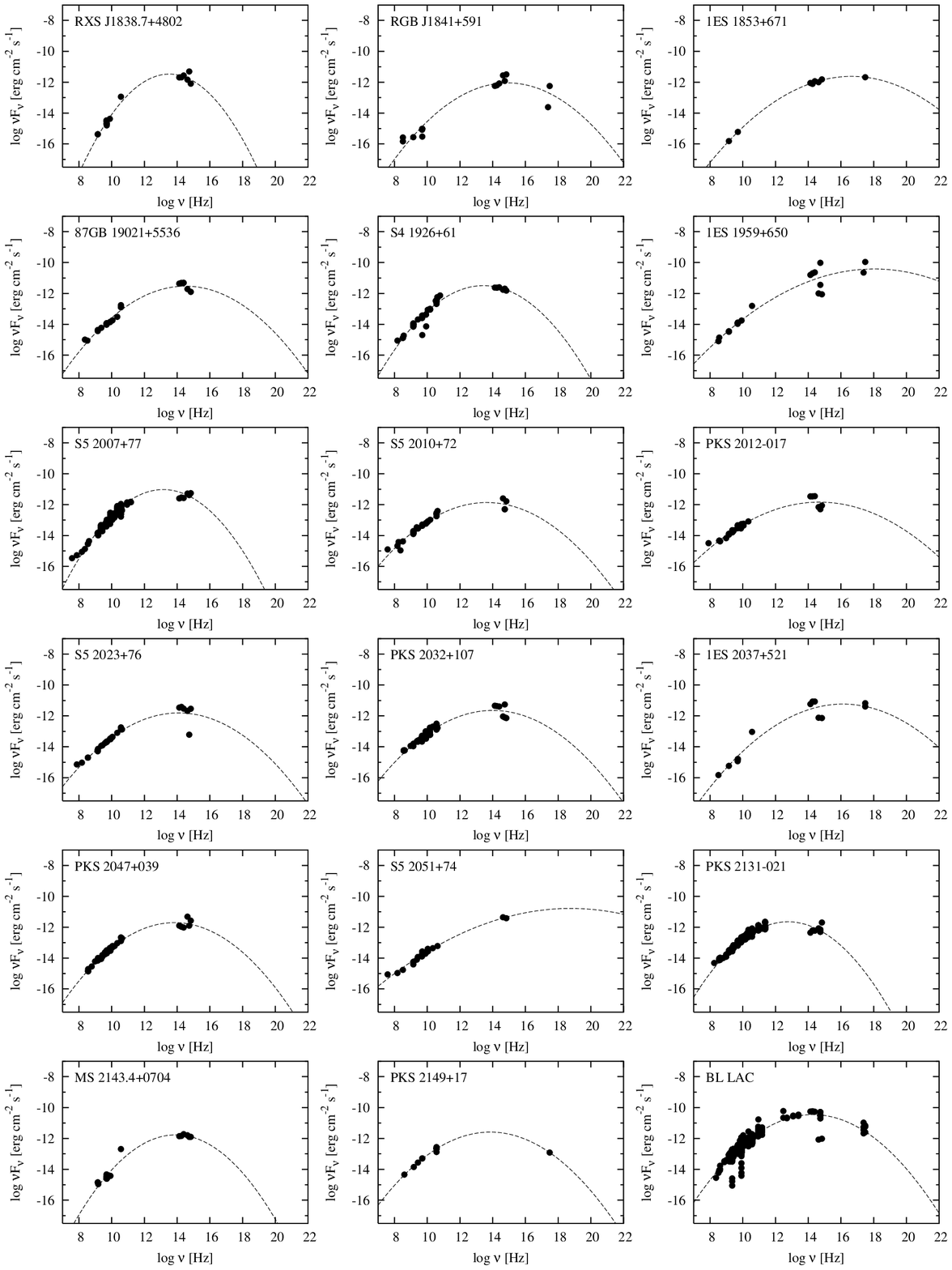}
\caption{continued.}
\label{sed16}
\end{figure*}

\addtocounter{figure}{-1}
\begin{figure*}
\centering
\includegraphics[width=17cm]{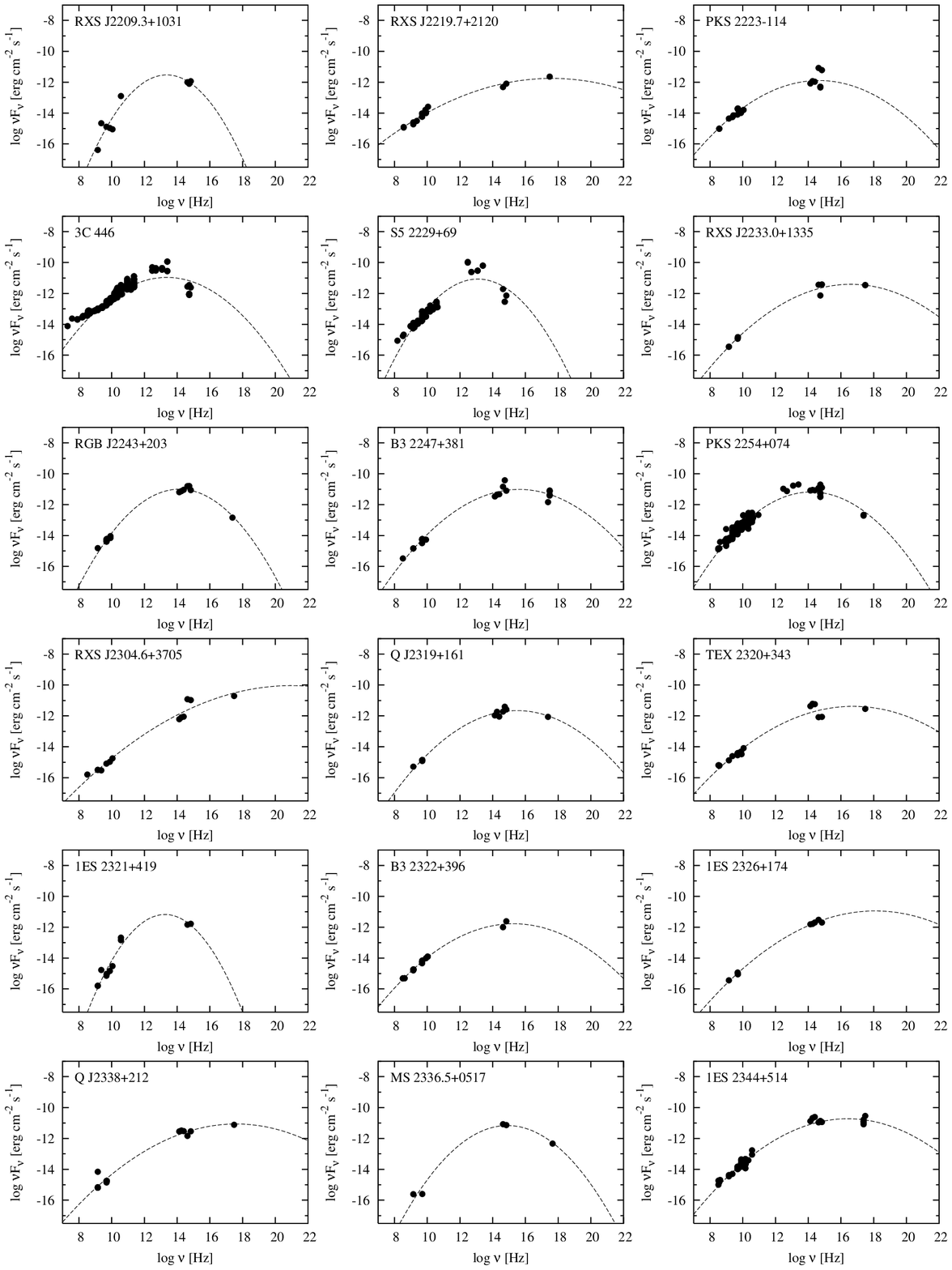}
\caption{continued.}
\label{sed17}
\end{figure*}

\addtocounter{figure}{-1}
\begin{figure*}
\centering
\includegraphics[width=17cm]{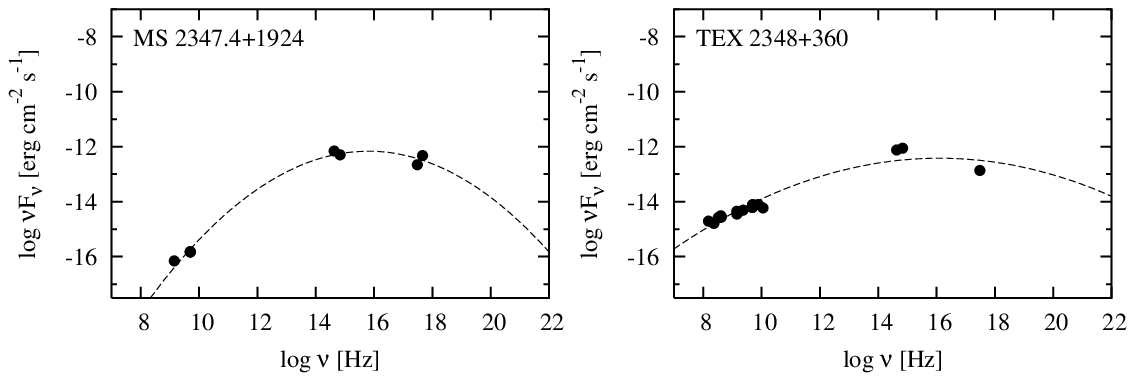}
\caption{continued.}
\label{sed18}
\end{figure*}

\end{document}